\documentclass[aps,preprint,amsfonts,amssymb,amsmath,eqsecnum,%
tightenlines,nofootinbib,showpacs,showkeys]{revtex4}
\newcommand{\cN}{\mathcal{N}}
\newcommand{\cP}{\mathcal{P}}
\newcommand{\cT}{\mathcal{T}}

\newcommand{\cV}{\mathcal{V}}

\newcommand{\bH}{\boldsymbol{H}}

\newcommand{\bQ}{\boldsymbol{Q}}

\newcommand{\fF}{\mathfrak{F}}

\newcommand{\bbN}{\mathbb{N}}
\newcommand{\bbZ}{\mathbb{Z}}

\newcommand{\rmd}{\mathrm{d}}%neIOPART
\newcommand{\del}{\partial}

\newcommand{\sV}{\mathsf{V}}
\begin{document}

%\begin{frontmatter}%ELSART

%\markboth{Authors' Names}
%{Instructions for Typing Manuscripts (Paper's Title)}%WSPC

%\catchline{}{}{}{}{}%WSPC
%
%%%%%%%%%%%%%%%%%%%%%%%%%%%%%%%%%%%%%%%%%%%%%%%%%%%%%%%%%%%%%%%%%%%%

\title{Parasupersymmetry and $\cN$-fold Supersymmetry in Quantum
 Many-Body Systems I. General Formalism and Second Order}
%\article[Short title]{TYPE}{Full title}%IOPART
%ELSART,WSPC
%\author{\corauthref{}},
%\corauth[]{}
%\ead{}%ELSART
%\author{Toshiaki Tanaka}
%\ead{ttanaka@mail.ncku.edu.tw}%ELSART
%\address{Department of Physics, National Cheng-Kung University,\\
% Tainan 701, Taiwan, R.O.C.}
%\address{National Center for Theoretical Sciences, Taiwan, R.O.C.}
%REVTEX4
\author{Toshiaki Tanaka}
\email{ttanaka@mail.ncku.edu.tw}
\affiliation{Department of Physics, National Cheng-Kung University,\\
 Tainan 701, Taiwan, R.O.C.\\
 National Center for Theoretical Sciences, Taiwan, R.O.C.}
%  \altaffiliation{}
%IOPART
%\author{Author 1\dag, Author 2\ddag and Toshiaki Tanaka\S}
%\address{\dag Address 1}
%\address{\ddag Address 2}
%\address{\S Department of Physics, National Cheng-Kung University,
% Tainan 701, Taiwan, R.O.C.}
%\address{\S National Center for Theoretical Sciences, Taiwan, R.O.C.}
%\eads{\mailto{email 1}, \mailto{email 2},
% \mailto{ttanaka@mail.ncku.edu.tw}}

%\date{\today}

\begin{abstract}

We propose an elegant formulation of parafermionic algebra and
parasupersymmetry of arbitrary order in quantum many-body systems
without recourse to any specific matrix representation of
parafermionic operators and any kind of deformed algebra.
Within our formulation, we show generically that every
parasupersymmetric quantum system of order $p$ consists of
$\cN$-fold supersymmetric pairs with $\cN\leq p$ and thus has
weak quasi-solvability and isospectral property. We also propose
a new type of non-linear supersymmetries, called quasi-parasupersymmetry,
which is less restrictive than parasupersymmetry and is different
from $\cN$-fold supersymmetry even in one-body systems though
the conserved charges are represented by higher-order linear
differential operators. To illustrate how our formulation works,
we construct second-order parafermionic algebra and three simple
examples of parasupersymmetric quantum systems of order 2, one is
essentially equivalent to the one-body Rubakov--Spiridonov type and
the others are two-body systems in which two supersymmetries are folded.
In particular, we show that the first model admits a generalized
$2$-fold superalgebra.

%\keywords{keyword1; keyword2; keyword3.}%WSPC
\end{abstract}

%\ccode{PACS Nos.: nn.nn.Xx; nn.nn.XX}%WSPC

%REVTEX4
%REVTEX4

%\begin{keyword}%ELSART
% parafermionic algebra\sep parasupersymmetry\sep $\cN$-fold
% supersymmetry\sep quasi-solvability
% \PACS 03.65.Fd\sep 11.30.Na\sep 11.30.Pb\sep 02.10.Hh
%\end{keyword}%ELSART

%\pacs{nn.nn.Xx, nn.nn.Xx, nn.nn.Xx, nn.nn.Xx}%IOPART
%\submitto{Journal}

%\preprint{XXX-xxx}%REVTEX4

%\end{frontmatter}%ELSART
\maketitle%neELSART

\section{Introduction}
\label{sec:intro}

Concept of symmetry has played a central role in the development of
modern theoretical physics and mathematical science. It may be almost
certain that there is an underlying symmetry if a system under
consideration exhibits a significant property that is not shared in
general cases. Thus, a discovery of a new symmetry enlarges our ability
and possibility to describe new phenomena both in the physical nature
and mathematical models. Even in the case when some physical requirements
prohibit any physically relevant model which has a certain kind of
symmetry, it can motivate us to consider another kind of symmetry.
For instance, the no-go theorem by Coleman and Mandula \cite{CM67},
which explains the failure of earlier attempts to unify the space-time
Poincar\'e symmetry and the approximate flavor symmetry within a larger
Lie algebra, promoted the study of supersymmetric theories. Although
the latter attempts arrived at another no-go theorem shown by
Haag, {\L}opusza\'{n}ski, and Sohnius \cite{HLS75}, so far there has been,
to the best of our knowledge, no other no-go theorems which can be
applicable to any kind of symmetry outside Lie superalgebra. Indeed,
this fact has motivated to investigate novel field theoretical models
with new kind of symmetry such as weak supersymmetry \cite{Sm03},
cubic supersymmetry \cite{MMT04,MTT04}, and so on.

Generally speaking, however, it is extremely difficult to construct
such a new quantum field theory in higher-dimensional space-time.
But if some new symmetry can be realized in higher-dimensional models,
we can always consider it in one space-time dimension, namely, in
quantum mechanical models. It means that if any quantum mechanical
systems cannot admit a certain symmetry, nor can quantum field
theoretical models (except for such a kind of symmetry which becomes
trivial only in one dimension). Hence, quantum mechanical models
provide a good touchstone to examine whether some symmetry has the
possibility to be realized in higher-dimensional theories. Furthermore,
they also provide a toy model to investigate non-trivial aspects
which one can hardly do in field theoretical models. In fact, the latter
was the reason why Witten introduced supersymmetric quantum mechanics
in order to acquire insight into non-perturbative aspects of
supersymmetric quantum field theory~\cite{Wi81,Wi82}. We note that
this strategy was also employed in the study of weak supersymmetry and
it was shown that some field theoretical models with weak supersymmetry
in one dimension reduces to $\cN$-fold supersymmetric quantum systems
with $\cN=2$~\cite{Sm03}.

The research field of $\cN$-fold supersymmetry in quantum mechanics was
initiated as a naive higher-derivative generalization of
the representation of supercharges~\cite{AIS93}. Later, its true
appreciation in connection with supersymmetric quantum field theory
was given by the proof of the equivalence to (weak)
quasi-solvability~\cite{AST01b}. That is, it was shown that
quasi-solvability, which is a less restrictive concept than quasi-exact
solvability~\cite{TU87,Us94}, is a one-dimensional analog and, in
a sense, a generalization of the perturbative non-renormalization
theorems in supersymmetric quantum field theory. In fact, this
connection naturally explains the disappearance of leading divergence
of perturbation series not only for the ground state but also for
a finite number of excited states of $\cN$-fold supersymmetric quantum
models, irrespective of whether $\cN$-fold supersymmetry is dynamically
broken or not~\cite{AKOSW99,ST02}. We note that it can provide
an implication in higher-dimensional theories; the connection between
weak and $\cN$-fold supersymmetries implies that characteristic aspects
of weak-supersymmetric quantum field theory, if exists, have much
intimate relation to those of $\cN$-fold supersymmetric quantum
mechanics. Thus, we can again recognize the importance of investigating
symmetry in quantum mechanical systems.

In this article, we would like to focus on parasupersymmetry. It is,
roughly speaking, symmetry between bosons and parafermions, and is
first proposed in one-body quantum mechanics~\cite{RS88}. However,
its characteristic features have been less understood than those of
$\cN$-fold supersymmetry. One of the reasons stems from the fact that
parasupersymmetric quantum mechanics has been usually formulated in
terms of matrix representations of parafermions~\cite{RS88,BD89a,BD89b,%
DV89,DV90a,BD90a,BD90b,DV90b,BD90c,BD90d,Mer90,BD91z,BD91a,BD91b,Mer91,%
AI91,SC91,Sp91,SV91,DMSV91,AISV91,To92,Kh92,Kh93,BDN93a,KMR93a,BDN93b,%
CKS95,BR95,BBKR97,JF98a,JF98b,JM99,Fa00,FJ00,MS00,SM01,BMQ02,CF03}.
Once a specific matrix representation is introduced, one can calculate
any product of parafermions which is not defined in the original
parafermionic algebra. As a result, it is difficult to appreciate which
part of the results is generic and which part of the results depends
on a specific choice of representations. Another reason is that
the mathematical relations among parasuperalgebras and various types
of deformed oscillator algebras including $q$-deformed ones have
attracted much more attention in the recent development of the research
field~\cite{FIK92,BD93a,De93b,IU93,Qu94,BDQ96,Pl97a,QV98,QV00,%
QV02,DK03,DK04} (see also the references cited in Refs.~\cite{QV00,DK03}).

Considering the situations described above,
we first propose an elegant formulation of parafermionic algebra
of arbitrary order without recourse to any specific representation
of parafermionic operators. It does not introduce any new type of
operators which do not exist in the fermionic case and does not
involve any kind of deformed algebra. With the aid of solely this
algebra, we then formulate parasupersymmetric quantum many-body systems
of arbitrary order in a generic way and then derive the general aspects
of parasupersymmetry. In particular, we show generically that every
parasupersymmetric quantum systems of order $p$ consists of $\cN$-fold
supersymmetric pairs with $\cN\leq p$. From this result, we
propose a generalization of parasupersymmetry which we would call
quasi-parasupersymmetry. Then we examine the second-order case in
detail.

We organize the article as follows. In the next section, we first
define parafermionic operators and a linear space on which they
act. Then, we propose postulates of parafermionic algebra. In
Section~\ref{sec:psusy}, after reviewing the definition of
parasupersymmetry, we formulate in a generic way parasupersymmetry
of arbitrary order in quantum many-body systems in terms of
the parafermionic operators just defined in Section~\ref{sec:palge}.
Using the parafermionic algebra, we derive the conditions for the
systems under consideration to have parasupersymmetry of arbitrary
order in terms of component parasupercharges and Hamiltonians.
From them we show that every parasupersymmetric system consists of
$\cN$-fold supersymmetric pairs with $\cN\leq p$ and thus has
weak quasi-solvability and isospectral property. Observing these
results, we propose a generalization of parasupersymmetry in
Section~\ref{sec:qpara}. Then, to see more explicitly how
the general formalism developed in Sections~\ref{sec:palge} and
\ref{sec:psusy} works, in Section~\ref{sec:palg2} we construct
systematically parafermionic algebra of order 2 solely based on
the postulates proposed in Section~\ref{sec:palge}. With the aid of
this parafermionic algebra of order 2, we investigate second-order
parasupersymmetric quantum systems and exhibit three simple examples
of such systems in Section~\ref{sec:2psqs}. Finally in the last
section, we summarize and discuss the results obtained in this paper
and possible future issues.

\section{Parafermionic Algebra}
\label{sec:palge}

First of all, let us define parafermionic algebra of order $p(\in\bbN)$.
It is an associative algebra composed of the identity operator $I$
and two parafermionic operators $\psi^{-}$ and $\psi^{+}$ of order
$p$ which satisfy the nilpotency:
\begin{align}
(\psi^{-})^{p}\neq 0,\quad (\psi^{+})^{p}\neq 0,\qquad
 (\psi^{-})^{p+1}=(\psi^{+})^{p+1}=0.
\label{eq:nilpo}
\end{align}
Hence, we immediately have $2p+1$ non-zero elements
$\{I,\psi^{-},\dots,(\psi^{-})^{p},\psi^{+},\dots,(\psi^{+})^{p}\}$.
We call them the \emph{fundamental} elements of parafermionic
algebra of order $p$. Parafermionic algebra is characterized by
anti-commutation relation $\{A,B\}=AB+BA$ and commutation relation
$[A,B]=AB-BA$ among the fundamental elements. When $p=1$, the algebra
must reduce to the ordinary fermionic one:
\begin{align}
\{\psi^{-},\psi^{-}\}=\{\psi^{+},\psi^{+}\}=0,\qquad
 \{\psi^{-},\psi^{+}\}=I.
\label{eq:falg}
\end{align}
The nilpotency (\ref{eq:nilpo}) for arbitrary $p$ is a trivial
generalization of the first fermionic algebra in Eq.~(\ref{eq:falg}).
Thus, the next problem is how to generalize the second relation
in Eq.~(\ref{eq:falg}).
In this paper, we propose the following relation as a generalization
of it to arbitrary order $p$: 
\begin{align}
\{\psi^{-},\psi^{+}\}+\{(\psi^{-})^{2},(\psi^{+})^{2}\}+\dots
 +\{(\psi^{-})^{p},(\psi^{+})^{p}\}=pI.
\label{eq:postu}
\end{align}
Before proceeding further with defining the algebra, we shall
next define parafermionic Fock spaces $\sV_{p}$ of order $p$ on
which the parafermionic operators act. The latter space is $(p+1)$
dimensional and its $p+1$ bases $|k\rangle$ ($k=0,\dots,p$)
are defined by
\begin{align}
\psi^{-}|0\rangle=0,\quad |k\rangle=(\psi^{+})^{k}|0\rangle,
 \quad\psi^{-}|k\rangle=|k-1\rangle\quad (k=1,\dots,p).
\label{eq:defpfs}
\end{align}
That is, $\psi^{-}$ and $\psi^{+}$ act as annihilation and
creation operators of parafermions, respectively.
The state $|0\rangle$ is called the parafermionic \emph{vacuum}.
The subspace spanned by each state $|k\rangle$ ($k=0,\dots,p$) is
called the $k$-parafermionic subspace and is denoted by
$\sV_{p}^{(k)}$. The adjoint vector $\langle k|$ of each $|k
\rangle$ is introduced as a linear operator which maps every
vector in $\sV_{p}$ into a complex number as follows:
\begin{align}
\langle k|l\rangle=\langle 0|(\psi^{-})^{k}|l\rangle,\quad
\langle 0|l\rangle=\langle 0|(\psi^{+})^{l}|0\rangle
 =\delta_{0,l}\quad (k,l=0,\dots,p).
\label{eq:defadv}
\end{align}
By the definitions (\ref{eq:defpfs}) and (\ref{eq:defadv}),
we immediately have a bi-orthogonal relation:
\begin{align}
\langle k|l\rangle=\delta_{k,l}\quad (k,l=0,\dots,p).
\end{align}
We can now define a set of projection operators
$\Pi_{k}:\sV_{p} \to\sV_{p}^{(k)}$ ($k=0,\dots,p$) which satisfy
\begin{align}
\Pi_{k}|l\rangle=\delta_{k,l}|k\rangle,\qquad\Pi_{k}\Pi_{l}=
 \delta_{k,l}\Pi_{k},\qquad\sum_{k=0}^{p}\Pi_{k}=I.
\label{eq:defpo}
\end{align}
When $p=1$, the projection operators are given in terms of the
fermionic operators by
\begin{align}
\Pi_{0}=\psi^{-}\psi^{+},\qquad\Pi_{1}=\psi^{+}\psi^{-}.
\label{eq:proj1}
\end{align}
At this stage, however, we do not know how the projection
operators are expressed in terms of the parafermionic operators
for arbitrary order $p>1$.
From the definitions (\ref{eq:defpfs}) and (\ref{eq:defpo}),
\begin{align*}
\Pi_{k}\psi^{+}|l\rangle&=\Pi_{k}|l+1\rangle
 =\delta_{k,l+1}|k\rangle,\\
\psi^{+}\Pi_{k}|l\rangle&=\delta_{k,l}\psi_{+}|k\rangle
 =\delta_{k,l}|k+1\rangle,
\end{align*}
and we obtain
\begin{align}
\Pi_{k+1}\psi^{+}=\psi^{+}\Pi_{k},
\label{eq:Pipsi+}
\end{align}
where and hereafter we put $\Pi_{k}\equiv 0$ for all $k<0$ and
$k>p$. Similarly, from the definitions (\ref{eq:defpfs}) and
(\ref{eq:defpo}), we obtain
\begin{align}
\psi^{-}\Pi_{k+1}=\Pi_{k}\psi^{-}.
\label{eq:psi-Pi}
\end{align}
We now come back to the parafermionic algebra. Apparently, the
relations (\ref{eq:nilpo}) and (\ref{eq:postu}) are not sufficient
for the determination of the full algebra. To determine other
multiplication relations we impose the following postulates:
\begin{enumerate}
\item First, the algebra must be consistent with
 Eq.~(\ref{eq:defpfs}).
 This requirement is indispensable for defining consistently
 the parafermionic Fock space $\sV_{p}$.
 When $p=1$, for instance, we have from the fermionic algebra
 (\ref{eq:falg})
 \begin{align*}
 \psi^{-}|1\rangle=\psi^{-}\psi^{+}|0\rangle
  =(I-\psi^{+}\psi^{-})|0\rangle=|0\rangle,
 \end{align*}
 thus the algebra is indeed consistent with Eq.~(\ref{eq:defpfs}).
\item Every projection operator $\Pi_{k}$ ($k=0,\dots,p$) can be
 expressed as a polynomial of the fundamental elements of the
 corresponding order $p$ so that the algebra is consistent with
 the definition (\ref{eq:defpo}).
\item Every product of three fundamental elements can be expressed
 as a polynomial of at most second-degree in the fundamental elements.
 These formulas are called the \emph{multiplication law}. For example,
 the multiplication law for the fermion operators ($p=1$) is given by
 \begin{align}
 \psi^{-}\psi^{+}\psi^{-}=\psi^{-},\qquad
  \psi^{+}\psi^{-}\psi^{+}=\psi^{+}.
 \label{eq:mult1}
 \end{align}
 We assume that the relations in Eq.~(\ref{eq:mult1}) hold for
 parafermionic operators of \emph{any} order $p$. As a consequence
 of this assumption, we immediately obtain for all $m,n\in\bbN$
 \begin{align}
 (\psi^{-})^{m}\psi^{+}(\psi^{-})^{n}=(\psi^{-})^{m+n-1},\quad
 (\psi^{+})^{m}\psi^{-}(\psi^{+})^{n}=(\psi^{+})^{m+n-1},
 \label{eq:mult2}
 \end{align}
 which also hold for arbitrary order.
\item We also assume that the following relations hold for arbitrary
 order:
 \begin{align}
 (\psi^{-})^{p}\psi^{+}\Pi_{p-1}=(\psi^{-})^{p-1}\Pi_{p-1},\qquad
 \psi^{+}(\psi^{-})^{p}\Pi_{p}=(\psi^{-})^{p-1}\Pi_{p}.
 \label{eq:mult3}
 \end{align}
 In the case of $p=1$, they are trivial; $\psi^{-}\psi^{+}\Pi_{0}
 =\Pi_{0}$ and $\psi^{+}\psi^{-}\Pi_{1}=\Pi_{1}$. As we will see later,
 the assumptions (\ref{eq:mult1}) and (\ref{eq:mult3}) turn to be
 crucial for a parasupersymmetric condition to make sense in quantum
 systems.
\end{enumerate}
We note that every polynomial composed of the fundamental elements
can be reduced to a polynomial of at most second-degree in them
as a consequence of the third postulate and the associativity.
Hence, together with the second postulate it means in particular
every projection operators must be expressed as a polynomial of
second-degree in the fundamental elements.

Finally, we introduce the quantity of parafermionic \emph{degree}
of operators as follows:
\begin{align}
\deg I=0,\qquad\deg\psi^{+}=1,\qquad\deg\psi^{-}=p,\\
\deg AB\equiv\deg A+\deg B\pmod{p+1}.
\end{align}
For example, $\deg(\psi^{+})^{k}=k$ and $\deg(\psi^{-})^{k}=p+1-k$
($k=1,\dots,p$).

\section{Parasupersymmetry}
\label{sec:psusy}

Parasupersymmetry of order 2 in quantum mechanics was first introduced
by Rubakov and Spiridonov \cite{RS88} and was later generalized
to arbitrary order independently by Tomiya~\cite{To92} and by
Khare~\cite{Kh92}. A different formulation for order 2 was proposed
by Beckers and Debergh \cite{BD90a} and a generalization of the latter
to arbitrary order was attempted by Chenaghlou and Fakhri \cite{CF03}.
Thus, we call them RSTK and BDCF formalism, respectively.
To define a $p$th-order parasupersymmetric system, we first introduce
a pair of parasupercharges $\bQ^{\pm}$ of order $p$ which satisfy
\begin{align}
(\bQ^{-})^{p}\neq 0,\quad(\bQ^{+})^{p}\neq 0,\quad
 (\bQ^{-})^{p+1}=(\bQ^{+})^{p+1}=0.
\label{eq:pfsc1}
\end{align}
A system $\bH$ is said to have \emph{parasupersymmetry of order $p$} if
it commutes with the parasupercharges of order $p$
\begin{align}
[\bQ^{-},\bH]=[\bQ^{+},\bH]=0,
\label{eq:pfsc2}
\end{align}
and satisfies the non-linear relations in the RSTK formalism
\begin{align}
\sum_{k=0}^{p}(\bQ^{-})^{p-k}\bQ^{+}(\bQ^{-})^{k}
 =C_{p}(\bQ^{-})^{p-1}\bH,\quad
\sum_{k=0}^{p}(\bQ^{+})^{p-k}\bQ^{-}(\bQ^{+})^{k}
 =C_{p}\bH(\bQ^{+})^{p-1},
\label{eq:pfsc3}
\end{align}
or in the BDCF formalism
\begin{subequations}
\label{eq:pfsc4}
\begin{align}
\underbrace{[\bQ^{-},\cdots,[\bQ^{-}}_{(p-1)\text{ times}},
 [\bQ^{+},\bQ^{-}]]\cdots]&=(-1)^{p}C_{p}(\bQ^{-})^{p-1}\bH,\\
\underbrace{[\bQ^{+},\cdots,[\bQ^{+}}_{(p-1)\text{ times}},
 [\bQ^{-},\bQ^{+}]]\cdots]&=C_{p}\bH(\bQ^{+})^{p-1},
\end{align}
\end{subequations}
where $C_{p}$ is a constant.
An apparent drawback of the BDCF formalism is that the relations
(\ref{eq:pfsc4}) do not reduce to the ordinary supersymmetric
anti-commutation relation $\{\bQ^{-},\bQ^{+}\}=C_{1}\bH$ when $p=1$,
in contrast to the RSTK relation (\ref{eq:pfsc3}). For this reason, we
discard the BDCF formalism in this paper though its defect may be
amended by, e.g., replacing all the commutators in (\ref{eq:pfsc4})
by anti-commutators, graded commutators $[A,B\}=AB-
(-1)^{\deg A\cdot\deg B}BA$, and so on.

An immediate consequence of the commutativity (\ref{eq:pfsc2}) is that
each $n$th-power of the parasupercharges ($2\leq n\leq p$) also commutes
with the system $\bH$
\begin{align}
[(\bQ^{-})^{n},\bH]=[(\bQ^{+})^{n},\bH]=0\quad(2\leq n\leq p).
\label{eq:pfsc5}
\end{align}
Hence, every parasupersymmetric system $\bH$ satisfying (\ref{eq:pfsc2})
always has $2p$ conserved charges.

To realize parasupersymmetry in quantum mechanical systems, we usually
consider a vector space $\fF\times\sV_{p}$ where $\fF$ is a linear space
of complex functions such as the Hilbert space $L^{2}$ in Hermitian
quantum theory and the Krein space $L_{\cP}^{2}$ in
$\cP\cT$-symmetric quantum theory \cite{Ta06b,Ta06d}. A parafermionic
quantum system $\bH$ is introduced by
\begin{align}
\bH=\sum_{k=0}^{p}H_{k}\Pi_{k},
\label{eq:pfqs}
\end{align}
where $H_{k}$ ($k=0,\dots,p$) are scalar Hamiltonians of $p$ variables
acting on $\fF$:
\begin{align}
H_{k}=-\frac{1}{2}\sum_{i=1}^{p}\frac{\partial_{i}^{2}}{\partial
 q_{i}^{2}}+V_{k}(q_{1},\cdots,q_{p})\quad (k=0,\dots,p).
\label{eq:Schro}
\end{align}
Two parasupercharges $\bQ^{\pm}$ are defined by
\begin{align}
\bQ^{-}=\sum_{k=0}^{p}Q_{k}^{-}\psi^{-}\Pi_{k},
 \qquad\bQ^{+}=\sum_{k=0}^{p}Q_{k}^{+}\Pi_{k}\psi^{+},
\label{eq:pfsc}
\end{align}
where $Q_{k}^{+}$ ($k=0,\dots,p$) are first-order linear operators
acting on $\fF$
\begin{align}
Q_{k}^{+}=\sum_{i=1}^{p}w_{k,i}(q_{1},\dots,q_{p})\frac{\partial}{
 \partial q_{i}}+W_{k}(q_{1},\dots,q_{p})\quad(k=0,\dots,p),
\label{eq:compQ}
\end{align}
and for each $k$, $Q_{k}^{-}$ is given by a certain `adjoint'
of $Q_{k}^{+}$, e.g., the (ordinary) adjoint $Q_{k}^{-}=
(Q_{k}^{+})^{\dagger}$ in the Hilbert space $L^{2}$,
the $\cP$-adjoint $Q_{k}^{-}=\cP (Q_{k}^{+})^{\dagger}\cP$ in
the Krein space $L_{\cP}^{2}$, and so on. For all $k\leq 0$ and
$k>p$ we put $Q_{k}^{\pm}\equiv 0$.
When $p=1$, the triple $(\bH,\bQ^{-},\bQ^{+})$ defined in
Eqs.~(\ref{eq:pfqs}) and (\ref{eq:pfsc}) becomes
\begin{align}
\bH&=\sum_{k=0}^{1}H_{k}\Pi_{k}=H_{0}\psi^{-}\psi^{+}
 +H_{1}\psi^{+}\psi^{-},\\
\bQ^{-}&=\sum_{k=0}^{1}Q_{k}^{-}\psi^{-}\Pi_{k}=Q_{1}^{-}\psi^{-},\\
\bQ^{+}&=\sum_{k=0}^{1}Q_{k}^{+}\Pi_{k}\psi^{+}=Q_{1}^{+}\psi^{+},
\end{align}
where Eqs.~(\ref{eq:proj1}) and (\ref{eq:mult1}) are used, and
thus reduces to an ordinary supersymmetric quantum mechanical
system \cite{Wi81}. The non-linear relation (\ref{eq:pfsc3}) together
with the nilpotency (\ref{eq:pfsc1}) for $p=1$ are just the
anti-commutation relations between supercharges
\begin{align}
\{\bQ^{\pm},\bQ^{\pm}\}=0,\qquad\{\bQ^{-},\bQ^{+}\}=C_{1}\bH.
\end{align}
Hence, the parasupersymmetric quantum systems defined by
Eqs.~(\ref{eq:pfsc1})--(\ref{eq:compQ}) provide a natural
generalization of ordinary supersymmetric quantum mechanics.

We next note that the parasupercharges $\bQ^{\pm}$ defined by
Eq.~(\ref{eq:pfsc}) already satisfy the nilpotency (\ref{eq:pfsc1}).
This is an immediate consequence of the following formulas:
\begin{align}
(\bQ^{-})^{n}&=\sum_{k=0}^{p}Q_{k-n+1}^{-}\cdots Q_{k-1}^{-}Q_{k}^{-}
 (\psi^{-})^{n}\Pi_{k}\quad (n\in\bbN),
\label{eq:Q-n}\\
(\bQ^{+})^{n}&=\sum_{k=0}^{p}Q_{k}^{+}Q_{k-1}^{+}\cdots Q_{k-n+1}^{+}
 \Pi_{k}(\psi^{+})^{n}\quad (n\in\bbN).
\label{eq:Q+n}
\end{align}
These formulas are easily proved by induction. We first note that
they are identical with the defining equations (\ref{eq:pfsc}) of
$\bQ^{\pm}$ when $n=1$. Suppose, for instance, Eq.~(\ref{eq:Q-n})
holds for a natural number $n(<p)$. Then, using Eqs.~(\ref{eq:defpo})
and (\ref{eq:psi-Pi}) we have,
\begin{align}
(\bQ^{-})^{n+1}&=\sum_{k,l=0}^{p}Q_{l-n+1}^{-}\cdots Q_{l-1}^{-}
 Q_{l}^{-}Q_{k}^{-}(\psi^{-})^{n}\Pi_{l}\psi^{-}\Pi_{k}\notag\\
&=\sum_{k,l=0}^{p}Q_{l-n+1}^{-}\cdots Q_{l-1}^{-}Q_{l}^{-}Q_{k}^{-}
 (\psi^{-})^{n+1}\Pi_{l+1}\Pi_{k}\notag\\
&=\sum_{k=0}^{p}Q_{k-n}^{-}\cdots Q_{k-2}^{-}Q_{k-1}^{-}Q_{k}^{-}
 (\psi^{-})^{n+1}\Pi_{k}.
\end{align}
The latter formula is nothing but Eq.~(\ref{eq:Q-n}) with $n$ replaced
by $n+1$. Thus we have completed the proof of the formula (\ref{eq:Q-n}).
In the same way, we can prove the other formula (\ref{eq:Q+n}).
Then, the nilpotency (\ref{eq:pfsc1}) immediately follows from
Eq.~(\ref{eq:nilpo}) and thus is always guaranteed by
the definition (\ref{eq:pfsc}).

The commutation relations of $\bH$ in (\ref{eq:pfqs}) and $\bQ^{\pm}$
in (\ref{eq:pfsc}) are calculated with the aid of Eqs.~(\ref{eq:defpo}),
(\ref{eq:Pipsi+}), and (\ref{eq:psi-Pi}) as
\begin{align}
[\bQ^{-},\bH]&=\sum_{k,l=0}^{p}Q_{k}^{-}H_{l}\psi^{-}\Pi_{k}\Pi_{l}
 -\sum_{k,l=0}^{p}H_{l}Q_{k}^{-}\Pi_{l}\psi^{-}\Pi_{k}\notag\\
&=\sum_{k=0}^{p}Q_{k}^{-}H_{k}\psi^{-}\Pi_{k}
 -\sum_{k=0}^{p}H_{k-1}Q_{k}^{-}\psi^{-}\Pi_{k},\\
[\bQ^{+},\bH]&=\sum_{k,l=0}^{p}Q_{k}^{+}H_{l}\Pi_{k}\psi^{+}\Pi_{l}
 -\sum_{k,l=0}^{p}H_{l}Q_{k}^{+}\Pi_{l}\Pi_{k}\psi^{+}\notag\\
&=\sum_{k=0}^{p}Q_{k}^{+}H_{k-1}\Pi_{k}\psi^{+}
 -\sum_{k=0}^{p}H_{k}Q_{k}^{+}\Pi_{k}\psi^{+}.
\end{align}
Hence, the commutativity (\ref{eq:pfsc2}) is satisfied if and only
if
\begin{align}
H_{k-1}Q_{k}^{-}=Q_{k}^{-}H_{k},\quad 
 Q_{k}^{+}H_{k-1}=H_{k}Q_{k}^{+},\quad\forall k=1,\dots,p.
\label{eq:inter}
\end{align}
That is, each pair of $H_{k-1}$ and $H_{k}$ must satisfy the
intertwining relations with respect to $Q_{k}^{-}$ and $Q_{k}^{+}$.
Furthermore, the commutation relations between $(\bQ^{\pm})^{n}$ and
$\bH$ ($2\leq n\leq p$) are similarly calculated by using
Eqs.~(\ref{eq:defpo}), (\ref{eq:Pipsi+}), (\ref{eq:psi-Pi}),
(\ref{eq:Q-n}), and (\ref{eq:Q+n}) as
\begin{align}
[(\bQ^{-})^{n},\bH]=&\,\sum_{k=0}^{p}Q_{k-n+1}^{-}\cdots Q_{k-1}^{-}
 Q_{k}^{-}H_{k}(\psi^{-})^{n}\Pi_{k}\notag\\
&\,-\sum_{k=0}^{p}H_{k-n}Q_{k-n+1}^{-}\cdots Q_{k-1}^{-}Q_{k}^{-}
 (\psi^{-})^{n}\Pi_{k},
\label{eq:Q-nH}\\
[(\bQ^{+})^{n},\bH]=&\,\sum_{k=0}^{p}Q_{k}^{+}Q_{k-1}^{+}\cdots
 Q_{k-n+1}^{+}H_{k-n}\Pi_{k}(\psi^{+})^{n}\notag\\
&\,-\sum_{k=0}^{p}H_{k}Q_{k}^{+}Q_{k-1}^{+}\cdots Q_{k-n+1}^{+}
 \Pi_{k}(\psi^{+})^{n}.
\end{align}
Hence, from the commutativity (\ref{eq:pfsc5}) any pair of $H_{k-n}$
and $H_{k}$ ($1\leq n\leq k\leq p$) satisfies
\begin{subequations}
\label{eq:Nfold}
\begin{align}
H_{k-n}Q_{k-n+1}^{-}\cdots Q_{k-1}^{-}Q_{k}^{-}
 =Q_{k-n+1}^{-}\dots Q_{k-1}^{-}Q_{k}^{-}H_{k},\\
Q_{k}^{+}Q_{k-1}^{+}\dots Q_{k-n+1}^{+}H_{k-n}
 =H_{k}Q_{k}^{+}Q_{k-1}^{+}\cdots Q_{k-n+1}^{+},
\end{align}
\end{subequations}
which means that $H_{k-n}$ and $H_{k}$ constitute a pair of
$\cN$-fold supersymmetry with $\cN=n$. The relations (\ref{eq:Nfold})
can be also derived by repeated applications of Eq.~(\ref{eq:inter}).
Since $\cN$-fold supersymmetry is essentially equivalent to weak
quasi-solvability \cite{AST01b,Ta03a}, parasupersymmetric quantum
systems also possess weak quasi-solvability. To see the structure
of weak quasi-solvability in the parasupersymmetric system $\bH$
more precisely, let us first define
\begin{align}
\cV_{n,k}^{-}=\ker (Q_{k-n+1}^{-}\cdots Q_{k}^{-}),\quad
 \cV_{n,k}^{+}=\ker (Q_{k}^{+}\cdots Q_{k-n+1}^{+})\quad
 (1\leq n\leq k\leq p).
\label{eq:defVnk}
\end{align}
By the definition (\ref{eq:defVnk}), the vector spaces
$\cV_{n,k}^{\pm}$ for each fixed $k$ are related as
\begin{align}
\cV_{1,k}^{-}\subset\cV_{2,k}^{-}\subset\cdots\subset
 \cV_{k,k}^{-},\quad
\cV_{1,k}^{+}\subset\cV_{2,k}^{+}\subset\cdots\subset
 \cV_{k,k}^{+},
\label{eq:flag}
\end{align}
On the other hand, it is evident from the intertwining relations
(\ref{eq:Nfold}) that each Hamiltonian $H_{k}$ ($0\leq k\leq p$)
preserves vector spaces as follows:
\begin{subequations}
\label{eq:invsp}
\begin{align}
H_{k}\cV_{n,k}^{-}\subset\cV_{n,k}^{-}\quad(1\leq n\leq k),\\
H_{k}\cV_{n,k+n}^{+}\subset\cV_{n,k+n}^{+}\quad(1\leq n\leq p-k).
\end{align}
\end{subequations}
From Eqs.~(\ref{eq:flag}) and (\ref{eq:invsp}), the largest
space preserved by each $H_{k}$ ($0\leq k\leq p$) is given by
\begin{align}
\cV_{k,k}^{-}+\cV_{p-k,p}^{+}\quad(0\leq k\leq p).
\label{eq:lstinv}
\end{align}
Needless to say, each Hamiltonian $H_{k}$ preserves the two spaces
in Eq.~(\ref{eq:lstinv}) separately.
The intertwining relations (\ref{eq:inter}) and (\ref{eq:Nfold})
ensure that all the component Hamiltonians $H_{k}$ ($k=0,\dots,p$) of
the system $\bH$ are isospectral outside the sectors $\cV_{n,k}^{\pm}$
($1\leq n\leq k\leq p$). The spectral degeneracy of $\bH$ in these
sectors depends on the form of each component of the parasupercharges,
$Q_{k}^{\pm}$ ($k=1,\dots,p$), and its structure can be very complicated
even in the case of second-order, see e.g. Refs.~\cite{Mo96a,Mo97}.

In addition to those `power-type' symmetries, every parasupersymmetric
quantum system $\bH$ defined in Eq.~(\ref{eq:pfqs}) can have
`discrete-type' ones. To see it, we first recall the basic fact that
for each operator $O$ which commutes with the system $\bH$, the
operator defined by $[O,\bQ^{\pm}]$ also commutes with $\bH$. It is
an immediate consequence of the Jacobi identity
\begin{align}
[[O,\bQ^{\pm}],\bH]=-[[\bH,O],\bQ^{\pm}]-[[\bQ^{\pm},\bH],O]=0.
\label{eq:Jacob}
\end{align}
Then, it follows from the intertwining relations (\ref{eq:Pipsi+})
and (\ref{eq:psi-Pi}) that every $p$th-order parasupersymmetric
quantum system $\bH$ commutes with $(\psi^{-})^{n}(\psi^{+})^{n}$
and $(\psi^{+})^{n}(\psi^{-})^{n}$ ($n=1,\cdots,p$), respectively;
for instance,
\begin{align}
[\bH,(\psi^{-})^{n}(\psi^{+})^{n}]&=\sum_{k=0}^{p}H_{k}\Pi_{k}
 (\psi^{-})^{n}(\psi^{+})^{n}-\sum_{k=0}^{p}H_{k}(\psi^{-})^{n}
 (\psi^{+})^{n}\Pi_{k}\notag\\
&=\sum_{k=0}^{p-n}H_{k}(\psi^{-})^{n}\Pi_{k+n}(\psi^{+})^{n}
 -\sum_{k=0}^{p-n}H_{k}(\psi^{-})^{n}\Pi_{k+n}(\psi^{+})^{n}=0.
\end{align}
Hence, if we define
\begin{align}
\bQ_{\{n\}}^{\pm}=[\{(\psi^{-})^{n},(\psi^{+})^{n}\},\bQ^{\pm}],
 \quad\bQ_{[n]}^{\pm}=[[(\psi^{-})^{n},(\psi^{+})^{n}],\bQ^{\pm}]
 \quad (n=1,\ldots,p),
\label{eq:discc}
\end{align}
all of $\bQ_{\{n\}}^{\pm}$ and $\bQ_{[n]}^{\pm}$ commute with $\bH$,
cf. Eq.~(\ref{eq:Jacob}):
\begin{align}
[\bQ_{\{n\}}^{\pm},\bH]=[\bQ_{[n]}^{\pm},\bH]=0\quad
 (n=1,\dots,p).
\end{align}
Explicitly, those conserved charges are expressed as
\begin{subequations}
\begin{align}
\bQ_{\{n\}}^{-}=\sum_{k=0}^{p}Q_{k}^{-}[\{ (\psi^{-})^{n},
 (\psi^{+})^{n}\},\psi^{-}]\Pi_{k},\quad\bQ_{[n]}^{-}
 =\sum_{k=0}^{p}Q_{k}^{-}[[(\psi^{-})^{n},(\psi^{+})^{n}],
 \psi^{-}]\Pi_{k},\\
\bQ_{\{n\}}^{+}=\sum_{k=0}^{p}Q_{k}^{+}\Pi_{k}[\{(\psi^{-})^{n},
 (\psi^{+})^{n}\},\psi^{+}],\quad\bQ_{[n]}^{+}=\sum_{k=0}^{p}
 Q_{k}^{+}\Pi_{k}[[(\psi^{-})^{n},(\psi^{+})^{n}],\psi^{+}].
\end{align}
\end{subequations}
We note, however, that they are in general not linearly independent
and we cannot determine the number of linearly independent conserved
charges without the knowledge of parafermionic algebra of each order.
For $n=1$, the conserved charges $\bQ_{\{1\}}^{\pm}$ and
$\bQ_{[1]}^{\pm}$ admit simpler expressions thanks to the assumption
(\ref{eq:mult1}):
\begin{subequations}
\label{eq:discc1}
\begin{align}
\bQ_{\{1\}}^{-}=\sum_{k=0}^{p}Q_{k}^{-}[\psi^{+},(\psi^{-})^{2}]
 \Pi_{k},\quad\bQ_{[1]}^{-}=2\bQ^{-}-\sum_{k=0}^{p}Q_{k}^{-}
 \{\psi^{+},(\psi^{-})^{2}\}\Pi_{k},\\
\bQ_{\{1\}}^{+}=\sum_{k=0}^{p}Q_{k}^{+}\Pi_{k}[\psi^{-},
 (\psi^{+})^{2}],\quad\bQ_{[1]}^{+}=\sum_{k=0}^{p}Q_{k}^{+}\Pi_{k}
 \{\psi^{-},(\psi^{+})^{2}\}-2\bQ^{+}.
\end{align}
\end{subequations}
From these expressions, we obtain for $p=1$
\begin{align}
\bQ_{\{1\}}^{\pm}=0,\qquad\bQ_{[1]}^{-}=2\bQ^{-},\qquad
 \bQ_{[1]}^{+}=-2\bQ^{+},
\end{align}
and thus there are no new conserved charges in supersymmetric
quantum mechanics.
As we will later show in Section~\ref{sec:2psqs}, in the case of
second-order parasupersymmetric quantum systems they exactly
corresponds to the additional charges and symmetry reported in
Ref.~\cite{DV90a}.

The non-linear relations (\ref{eq:pfsc3}) can be also calculated in
a similar way. Using Eqs.~(\ref{eq:defpo}), (\ref{eq:Pipsi+}),
(\ref{eq:psi-Pi}), (\ref{eq:pfqs}), (\ref{eq:pfsc}), and
(\ref{eq:Q-n}), we obtain
\begin{align}
(\bQ^{-})^{p}\bQ_{+}=&\, Q_{1}^{-}\cdots Q_{p}^{-}Q_{p}^{+}
 (\psi^{-})^{p}\psi^{+}\Pi_{p-1},
\label{eq:lhsnl}\\
(\bQ^{-})^{p-k}\bQ^{+}(\bQ^{-})^{k}=&\,\sum_{l=p-1}^{p}Q_{l-p+2}^{-}
 \cdots Q_{l-k+1}^{-}Q_{l-k+1}^{+}\notag\\
&\,\times Q_{l-k+1}^{-}\cdots Q_{l}^{-}(\psi^{-})^{p-k}\psi^{+}
 (\psi^{-})^{k}\Pi_{l}\quad (1\leq k\leq p-1),\\[10pt]
\bQ^{+}(\bQ^{-})^{p}=&\, Q_{1}^{+}Q_{1}^{-}\cdots Q_{p}^{-}
 \psi^{+}(\psi^{-})^{p}\Pi_{p},\\
(\bQ^{-})^{p-1}\bH=&\,\sum_{l=p-1}^{p}Q_{l-p+2}^{-}\cdots Q_{l}^{-}
 H_{l}(\psi^{-})^{p-1}\Pi_{l}\quad (p\geq 2).
\label{eq:rhsnl}
\end{align}
Employing the assumptions (\ref{eq:mult2}) and (\ref{eq:mult3}), we
conclude that the first non-linear relation in Eq.~(\ref{eq:pfsc3})
is satisfied if and only if the following two identities hold:
\begin{subequations}
\label{eq:parac}
\begin{align}
Q_{1}^{-}\cdots Q_{p}^{-}Q_{p}^{+}+\sum_{k=1}^{p-1}Q_{1}^{-}
 \cdots Q_{p-k}^{-}Q_{p-k}^{+}Q_{p-k}^{-}\cdots Q_{p-1}^{-}
 =C_{p}Q_{1}^{-}\cdots Q_{p-1}^{-}H_{p-1},
\label{eq:parac1}\\
\sum_{k=1}^{p-1}Q_{2}^{-}\cdots Q_{p-k+1}^{-}Q_{p-k+1}^{+}
 Q_{p-k+1}^{-}\cdots Q_{p}^{-}+Q_{1}^{+}Q_{1}^{-}\cdots Q_{p}^{-}
 =C_{p}Q_{2}^{-}\cdots Q_{p}^{-}H_{p}.
\end{align}
\end{subequations}
The conditions for the second non-linear relation in
Eq.~(\ref{eq:pfsc3}) are apparently given by the `adjoint' of
Eqs.~(\ref{eq:parac}). Here we see the crucial role played by the
formulas (\ref{eq:mult2}) and (\ref{eq:mult3}): without them,
the non-linear condition (\ref{eq:pfsc3}) would just lead to
the conclusion that every component appeared in
Eqs.~(\ref{eq:lhsnl})--(\ref{eq:rhsnl}) must vanish separately,
which would result in the trivial system $\bQ^{\pm}=\bH=0$.
Hence, the assumptions (\ref{eq:mult1}) and (\ref{eq:mult3}) are
indispensable for defining non-trivial parasupersymmetric quantum
systems.

\section{Quasi-parasupersymmetry}
\label{sec:qpara}

In the previous section, we have found that every pair of the
component Hamiltonians possesses $\cN$-fold supersymmetry with
$\cN\leq p$ and thus has isospectral property and weak
quasi-solvability. However, $\cN$-fold supersymmetric systems
constructed from $\cN$ repeated applications of first-order
intertwining relations as in the present case (cf.
Eq.~(\ref{eq:inter})) are well known to be quite restrictive
in comparison with general $\cN$-fold supersymmetric systems
(see Ref.~\cite{AST01a} in the case of type A $\cN$-fold
supersymmetry). This naturally explains the fact that almost
all the parasupersymmetric potentials so far found
are shape-invariant types (see e.g.
Refs.~\cite{BD91z,BDN93a,CKS95,JF98a}).
In this sense, we can say the conditions of parasupersymmetry
(\ref{eq:pfsc1})--(\ref{eq:pfsc3}) are too strong to obtain
various non-trivial models having several intriguing physical
consequences.

The above observation, if we take into account the fact that
the strict first-order intertwining relations (\ref{eq:inter})
directly come from the commutation relation (\ref{eq:pfsc2}),
leads us to consider a less restrictive condition as follows.
With a given pair of parasupercharges $\bQ^{\pm}$ of order $p$
which satisfy the nilpotency (\ref{eq:pfsc1}), a system $\bH$
is said to have \emph{quasi-parasupersymmetry of order} $(p,q)$
if there exists a natural number $q$ ($1\leq q\leq p$) such
that $(\bQ^{\pm})^{q}$ commutes with $\bH$ and the non-linear
constraint (\ref{eq:pfsc3}) is satisfied. That is, it is
characterized by the following algebraic relations:
\begin{align}
(\bQ^{-})^{p}\neq 0,\quad (\bQ^{+})^{p}\neq 0,\quad
 (\bQ^{-})^{p+1}=(\bQ^{+})^{p+1}=0,\\[10pt]
[(\bQ^{-})^{q},\bH]=[(\bQ^{+})^{q},\bH]=0\quad (1\leq q\leq p),\\
\sum_{k=0}^{p}(\bQ^{-})^{p-k}\bQ^{+}(\bQ^{-})^{k}
 =C_{p}(\bQ^{-})^{p-1}\bH,\quad
\sum_{k=0}^{p}(\bQ^{+})^{p-k}\bQ^{-}(\bQ^{+})^{k}
 =C_{p}\bH(\bQ^{+})^{p-1}.
\label{eq:qpfsc3}
\end{align}
By definition, quasi-parasupersymmetry of order $(p,q)$ reduces
to (ordinary) parasupersymmetry when $q=1$. Thus, it can be
regarded as a generalization of parasupersymmetry. A key
ingredient of this new symmetry is that the commutativity
$[(\bQ^{\pm})^{n},\bH]=0$ for $n<q$ is not necessarily
fulfilled in contrast to (ordinary) parasupersymmetry.
As a consequence, only the less restrictive $q$th-order intertwining
relations (\ref{eq:Nfold}) with $n=q$ should be satisfied between
every pair of $H_{k-q}$ and $H_{k}$ in the case of
quasi-parasupersymmetry of order $(p,q)$. The `power-type' conserved
charges in this case are apparently given by
\begin{align}
[(\bQ^{-})^{qn},\bH]=[(\bQ^{+})^{qn},\bH]=0\quad (2\leq n\leq
 [{\textstyle\frac{p}{q}}]),
\end{align}
where $[x]$ is the maximum integer which does not exceed $x$,
and thus the number of conserved charges is reduced to
$2[\frac{p}{q}]$. It is evident that parasupersymmetry of order
$p$ always implies quasi-parasupersymmetry of order $(p,q)$ for
all $q=1,\dots,p$. We note, however, that quasi-parasupersymmetry
of order $(p,q)$ does not necessarily imply that of order $(p,q+n)$
with $n>0$. This fact can be easily understood from the difference
of the conserved charges. For examples, the conserved charges
in order $(p,2)$ are $(\bQ^{\pm})^{2}$, $(\bQ^{\pm})^{4}$,
$\ldots$ while those in order $(p,3)$ are $(\bQ^{\pm})^{3}$,
$(\bQ^{\pm})^{6}$, $\ldots$, and thus they are different symmetries.
This observation clearly indicates that for every pair of natural
numbers $q_{1}$ and $q_{2}$ ($q_{1},q_{2}\leq p$)
quasi-parasupersymmetries of order $(p,q_{1})$ and $(p,q_{2})$ have
common conserved charges $(\bQ^{\pm})^{n}$ only when $n(\leq p)$ is
a common multiple of $q_{1}$ and $q_{2}$. The `discrete-type'
conserved charges (cf. Eq.~(\ref{eq:discc})) are similarly defined.

We note that the components of conserved charges of this symmetry
are given by higher-derivative linear differential operators and thus
it looks like a kind of $\cN$-fold supersymmetry. In particular,
the full algebra of quasi-parasupersymmetry of order $(p,p)$ includes
\begin{align}
\{(\bQ^{-})^{p},(\bQ^{-})^{p}\}=\{(\bQ^{+})^{p},(\bQ^{+})^{p}\}=0,
 \quad [(\bQ^{\pm})^{p},\bH]=0,
\end{align}
where the components of $(\bQ^{\pm})^{p}$ are $p$th-order
linear differential operators, and thus it resembles $\cN$-fold
supersymmetry with $\cN=p$. Indeed, in the case of a single
variable the anti-commutator $\{(\bQ^{-})^{p},(\bQ^{+})^{p}\}$ can
be expressed as a polynomial of degree $p$ in $\bH$ \cite{AST01b,AS03}
and thus the system exactly coincides with one-body $\cN$-fold
supersymmetry with $\cN=p$ \emph{if} the system admits
$H_{1}=\cdots=H_{p-1}\equiv 0$. However, the latter cannot be
the case in quasi-parasupersymmetry due to the first constraint
on $H_{p-1}$ in Eq.~(\ref{eq:parac1}), which comes from the
non-linear constraint (\ref{eq:qpfsc3}), unless the l.h.s. of
Eq.~(\ref{eq:parac1}) accidentally vanishes. Hence, despite
the resemblance of algebras and the fact that the conserved
charges are represented by $\cN$th-order linear differential
operators with $\cN=p$, quasi-parasupersymmetry provides a new
type of non-linear supersymmetries which is different from
$\cN$-fold supersymmetry even in one-body systems.

\section{Parafermionic Algebra of Order $2$}
\label{sec:palg2}

In this section, we shall construct parafermionic algebra of order 2
based on the postulates in Section~\ref{sec:palge}. The starting
point is the relations (\ref{eq:nilpo}) and (\ref{eq:postu}) for $p=2$:
\begin{align}
(\psi^{-})^{3}=(\psi^{+})^{3}=0,
\label{eq:2alg0}\\
\{\psi^{-},\psi^{+}\}+\{(\psi^{-})^{2},(\psi^{+})^{2}\}=2I.
\label{eq:2alg1}
\end{align}
First, multiplying (\ref{eq:2alg1}) by two $\psi^{-}$s as $(\psi^{-})^{2}
\times$(\ref{eq:2alg1}), $\psi^{-}\times$(\ref{eq:2alg1})$\times\psi^{-}$,
and (\ref{eq:2alg1})$\times(\psi^{-})^{2}$, and applying the nilpotency
(\ref{eq:2alg0}), we have
\begin{subequations}
\begin{align}
(\psi^{-})^{2}\psi^{+}\psi^{-}+(\psi^{-})^{2}(\psi^{+})^{2}(\psi^{-})^{2}
 &=2(\psi^{-})^{2},\\
(\psi^{-})^{2}\psi^{+}\psi^{-}+\psi^{-}\psi^{+}(\psi^{-})^{2}
 &=2(\psi^{-})^{2},\\
\psi^{-}\psi^{+}(\psi^{-})^{2}+(\psi^{-})^{2}(\psi^{+})^{2}(\psi^{-})^{2}
 &=2(\psi^{-})^{2}.
\end{align}
\end{subequations}
From the above set of equations, we immediately obtain
\begin{align}
(\psi^{-})^{2}\psi^{+}\psi^{-}=\psi^{-}\psi^{+}(\psi^{-})^{2}
 =(\psi^{-})^{2}(\psi^{+})^{2}(\psi^{-})^{2}=(\psi^{-})^{2},
\label{eq:2alg2}\\
\qquad(\psi^{-})^{2}\psi^{+}(\psi^{-})^{2}=0.
\label{eq:2alg2'}
\end{align}
We note that these formulas are consistent with the assumption
(\ref{eq:mult2}).
Next, multiplying (\ref{eq:2alg1}) by $\psi^{+}$ and $\psi^{-}$ as
$\psi^{+}\psi^{-}\times$(\ref{eq:2alg1}),
$\psi^{+}\times$(\ref{eq:2alg1})$\times\psi^{-}$, and
(\ref{eq:2alg1})$\times\psi^{+}\psi^{-}$, and applying the nilpotency
(\ref{eq:2alg0}) and the formula (\ref{eq:2alg2}), we have
\begin{subequations}
\begin{align}
\psi^{+}(\psi^{-})^{2}\psi^{+}+\psi^{+}\psi^{-}\psi^{+}\psi^{-}
 +(\psi^{+})^{2}(\psi^{-})^{2}&=2\psi^{+}\psi^{-},\\
\psi^{+}\psi^{-}\psi^{+}\psi^{-}+(\psi^{+})^{2}(\psi^{-})^{2}
 +\psi^{+}(\psi^{-})^{2}(\psi^{+})^{2}\psi^{-}&=2\psi^{+}\psi^{-},\\
\psi^{-}(\psi^{+})^{2}\psi^{-}+\psi^{+}\psi^{-}\psi^{+}\psi^{-}
 +(\psi^{+})^{2}(\psi^{-})^{2}&=2\psi^{+}\psi^{-}.
\end{align}
\end{subequations}
From the above set of equations, we obtain
\begin{align}
\psi^{+}(\psi^{-})^{2}\psi^{+}=\psi^{+}(\psi^{-})^{2}(\psi^{+})^{2}
 \psi^{-}=\psi^{-}(\psi^{+})^{2}\psi^{-}.
\label{eq:2alg3}
\end{align}
Multiplying (\ref{eq:2alg3}) by $\psi^{-}$ from left or right, and
applying the formula (\ref{eq:2alg2}), we get
\begin{align}
(\psi^{-})^{2}(\psi^{+})^{2}\psi^{-}=(\psi^{-})^{2}\psi^{+},\qquad
 \psi^{-}(\psi^{+})^{2}(\psi^{-})^{2}=\psi^{+}(\psi^{-})^{2}.
\label{eq:2alg4}
\end{align}
Next, we multiply Eq.~(\ref{eq:2alg1}) by $\psi^{-}$ from left to
have
\begin{align}
(\psi^{-})^{2}\psi^{+}+\psi^{-}\psi^{+}\psi^{-}+\psi^{-}
(\psi^{+})^{2}(\psi^{-})^{2}=2\psi^{-}.
\label{eq:2alg5}
\end{align}
Comparing Eqs.~(\ref{eq:2alg4}) and (\ref{eq:2alg5}), and using
the assumption (\ref{eq:mult1}), we get
\begin{align}
\{\psi^{+},(\psi^{-})^{2}\}=\psi^{-}\psi^{+}\psi^{-}=\psi^{-}.
\label{eq:2alg6}
\end{align}
Finally, multiplying the second formula in Eq.~(\ref{eq:2alg6})
by $\psi^{+}$ from left and right, we have
\begin{align}
(\psi^{+})^{2}(\psi^{-})^{2}+\psi^{+}(\psi^{-})^{2}\psi^{+}
 =\psi^{+}\psi^{-},\qquad\psi^{+}(\psi^{-})^{2}\psi^{+}
 +(\psi^{-})^{2}(\psi^{+})^{2}=\psi^{-}\psi^{+}.
\end{align}
Thus, we obtain the following formula:
\begin{align}
\psi^{+}(\psi^{-})^{2}\psi^{+}=\psi^{+}\psi^{-}-(\psi^{+})^{2}
 (\psi^{-})^{2}=\psi^{-}\psi^{+}-(\psi^{-})^{2}(\psi^{+})^{2}.
\label{eq:2alg7}
\end{align}
The second equality in Eq.~(\ref{eq:2alg7}) can be expressed as
\begin{align}
[\psi^{-},\psi^{+}]=[(\psi^{-})^{2},(\psi^{+})^{2}].
\end{align}
We note that all the formulas so far derived also hold when all
the indices of $+$ and $-$ are interchanged since the original
algebra (\ref{eq:2alg0}) and (\ref{eq:2alg1}) is invariant
under the interchange of $+$ and $-$. Thus, Eqs.~(\ref{eq:2alg2}),
(\ref{eq:2alg2'}), (\ref{eq:2alg4}), (\ref{eq:2alg6}), and
(\ref{eq:2alg7}) respectively imply
\begin{align}
(\psi^{+})^{2}\psi^{-}\psi^{+}=\psi^{+}\psi^{-}(\psi^{+})^{2}
 =(\psi^{+})^{2}(\psi^{-})^{2}(\psi^{+})^{2}=(\psi^{+})^{2},
\label{eq:2alg8}\\
(\psi^{+})^{2}\psi^{-}(\psi^{+})^{2}=0,\\
(\psi^{+})^{2}(\psi^{-})^{2}\psi^{+}=(\psi^{+})^{2}\psi^{-},
 \qquad\psi^{+}(\psi^{-})^{2}(\psi^{+})^{2}=\psi^{-}
 (\psi^{+})^{2},\\
\{\psi^{-}, (\psi^{+})^{2}\}=\psi^{+}\psi^{-}\psi^{+}
 =\psi^{+},\\
\psi^{-}(\psi^{+})^{2}\psi^{-}=\psi^{-}\psi^{+}-(\psi^{-})^{2}
 (\psi^{+})^{2}=\psi^{+}\psi^{-}-(\psi^{+})^{2}(\psi^{-})^{2}.
\label{eq:2alg11}
\end{align}
The fourth formula can be obtained also from Eq.~(\ref{eq:2alg3}).
Summarizing the results, we have derived second-order
parafermionic algebra as follows:
\begin{align}
(\psi^{-})^{3}=(\psi^{+})^{3}=0,
\label{eq:alg2a}\\
\{\psi^{-},(\psi^{+})^{2}\}=\psi^{+},\qquad
 \{\psi^{+},(\psi^{-})^{2}\}=\psi^{-},
\label{eq:alg2b}\\
\{\psi^{-},\psi^{+}\}+\{(\psi^{-})^{2},(\psi^{+})^{2}\}=2I,
\label{eq:alg2c}\\
[\psi^{-},\psi^{+}]=[(\psi^{-})^{2},(\psi^{+})^{2}].
\label{eq:alg2d}
\end{align}
The last two relations (\ref{eq:alg2c}) and (\ref{eq:alg2d})
together imply an important formula
\begin{align}
\psi^{-}\psi^{+}+(\psi^{+})^{2}(\psi^{-})^{2}
 =\psi^{+}\psi^{-}+(\psi^{-})^{2}(\psi^{+})^{2}=I.
\label{eq:alg2e}
\end{align}
Classifying the formulas (\ref{eq:2alg2}), (\ref{eq:2alg4}),
(\ref{eq:2alg6}), (\ref{eq:2alg7}), and
(\ref{eq:2alg8})--(\ref{eq:2alg11}) with respect to
parafermionic degrees, we have the following multiplication
law for the parafermionic operators of order 2:
\begin{itemize}
\item Degree 0:
 \begin{align*}
 (\psi^{-})^{2}\psi^{+}(\psi^{-})^{2}&=0,
  & \psi^{+}(\psi^{-})^{2}\psi^{+}&=\Pi_{1},\\
 \psi^{-}(\psi^{+})^{2}\psi^{-}&=\Pi_{1},
  & (\psi^{+})^{2}\psi^{-}(\psi^{+})^{2}&=0.
 \end{align*}
\item Degree 1:
 \begin{align*}
 (\psi^{-})^{2}\psi^{+}\psi^{-}&=(\psi^{-})^{2},
  & \psi^{+}(\psi^{-})^{2}(\psi^{+})^{2}&=\psi^{-}
  (\psi^{+})^{2},\\
 (\psi^{-})^{2}(\psi^{+})^{2}(\psi^{-})^{2}&=(\psi^{-})^{2},
  & \psi^{+}\psi^{-}\psi^{+}&=\psi^{+},\\
 \psi^{-}\psi^{+}(\psi^{-})^{2}&=(\psi^{-})^{2},
  & (\psi^{+})^{2}(\psi^{-})^{2}\psi^{+}&=(\psi^{+})^{2}
  \psi^{-}.
 \end{align*}
\item Degree 2:
 \begin{align*}
 (\psi^{-})^{2}(\psi^{+})^{2}\psi^{-}&=(\psi^{-})^{2}\psi^{+},
  & \psi^{+}\psi^{-}(\psi^{+})^{2}&=(\psi^{+})^{2},\\
 \psi^{-}\psi^{+}\psi^{-}&=\psi^{-},
  & (\psi^{+})^{2}(\psi^{-})^{2}(\psi^{+})^{2}
  &=(\psi^{+})^{2},\\
 \psi^{-}(\psi^{+})^{2}(\psi^{-})^{2}&=\psi^{+}(\psi^{-})^{2},
  & (\psi^{+})^{2}\psi^{-}\psi^{+}&=(\psi^{+})^{2}.
 \end{align*}
\end{itemize}
In the above, $\Pi_{1}$ is defined by
\begin{align}
\Pi_{1}=\psi^{+}\psi^{-}-(\psi^{+})^{2}(\psi^{-})^{2}
 =\psi^{-}\psi^{+}-(\psi^{-})^{2}(\psi^{+})^{2}.
\label{eq:2pjo1}
\end{align}
As we shall show in what follows, it is indeed the projection
operator into the 1-parafermionic subspace. From
the definition (\ref{eq:defpfs}) and the multiplication law,
\begin{align*}
\Pi_{1}|0\rangle&=(\psi^{+}\psi^{-}-(\psi^{+})^{2}
 (\psi^{-})^{2})|0\rangle=0,\\
\Pi_{1}|1\rangle&=(\psi^{+}\psi^{-}\psi^{+}-(\psi^{+})^{2}
 (\psi^{-})^{2}\psi^{+})|0\rangle=(\psi^{+}-(\psi^{+})^{2}
 \psi^{-})|0\rangle=|1\rangle,\\
\Pi_{1}|2\rangle&=(\psi^{+}\psi^{-}(\psi^{+})^{2}-(\psi^{+})^{2}
 (\psi^{-})^{2}(\psi^{+})^{2})|0\rangle=((\psi^{+})^{2}
 -(\psi^{+})^{2})|0\rangle=0.
\end{align*}
Similarly, the other projection operators $\Pi_{0}$ and $\Pi_{2}$
are given by
\begin{align}
\Pi_{0}=(\psi^{-})^{2}(\psi^{+})^{2},\qquad
 \Pi_{2}=(\psi^{+})^{2}(\psi^{-})^{2}.
\label{eq:2pjo2}
\end{align}
We can easily check with the aid of the multiplication law and
the formula (\ref{eq:alg2e}) that the operators $\Pi_{i}$ ($i=0,1,2$)
in Eqs.~(\ref{eq:2pjo1}) and (\ref{eq:2pjo2}) satisfy
the definition (\ref{eq:defpo}) for $p=2$. For example,
\begin{align*}
\Pi_{1}^{2}=&\,\psi^{+}\psi^{-}\psi^{+}\psi^{-}-\psi^{+}\psi^{-}
 (\psi^{+})^{2}(\psi^{-})^{2}\\
&\,-(\psi^{+})^{2}(\psi^{-})^{2}\psi^{+}\psi^{-}+(\psi^{+})^{2}
 (\psi^{-})^{2}(\psi^{+})^{2}(\psi^{-})^{2}\\
=&\,\psi^{+}\psi^{-}-(\psi^{+})^{2}(\psi^{-})^{2}-(\psi^{+})^{2}
 (\psi^{-})^{2}+(\psi^{+})^{2}(\psi^{-})^{2}=\Pi_{1},\\
\Pi_{1}\Pi_{2}=&\,\psi^{+}\psi^{-}(\psi^{+})^{2}(\psi^{-})^{2}
 -(\psi^{+})^{2}(\psi^{-})^{2}(\psi^{+})^{2}(\psi^{-})^{2}\\
=&\,(\psi^{+})^{2}(\psi^{-})^{2}-(\psi^{+})^{2}(\psi^{-})^{2}=0,
\end{align*}
and so on. The intertwining relations  (\ref{eq:Pipsi+}) and
(\ref{eq:psi-Pi}) can be easily checked as
\begin{align*}
\Pi_{1}\psi^{+}=\psi^{+}\Pi_{0}=\psi^{-}(\psi^{+})^{2},\qquad
 \Pi_{2}\psi^{+}=\psi^{+}\Pi_{1}=(\psi^{+})^{2}\psi^{-},\\
\psi^{-}\Pi_{1}=\Pi_{0}\psi^{-}=(\psi^{-})^{2}\psi^{+},\qquad
 \psi^{-}\Pi_{2}=\Pi_{1}\psi^{-}=\psi^{+}(\psi^{-})^{2}.
\end{align*}
We also note that the second-order parasuperalgebra
(\ref{eq:alg2a})--(\ref{eq:alg2e}) is consistent with
Eq.~(\ref{eq:defpfs}). From the relations (\ref{eq:alg2e})
and (\ref{eq:alg2b}), we have
\begin{align*}
\psi^{-}|1\rangle&=\psi^{-}\psi^{+}|0\rangle
 =(I-(\psi^{+})^{2}(\psi^{-})^{2})|0\rangle=|0\rangle,\\
\psi^{-}|2\rangle&=\psi^{-}(\psi^{+})^{2}|0\rangle
 =(\psi^{+}-(\psi^{+})^{2}\psi^{-})|0\rangle=|1\rangle,
\end{align*}
which are exactly Eq.~(\ref{eq:defpfs}). The assumption
(\ref{eq:mult3}) for $p=2$ is also satisfied as
\begin{align*}
(\psi^{-})^{2}\psi^{+}\Pi_{1}=\psi^{-}\Pi_{1}=(\psi^{-})^{2}
 \psi^{+},\qquad\psi^{+}(\psi^{-})^{2}\Pi_{2}=\psi^{-}\Pi_{2}
 =\psi^{+}(\psi^{-})^{2}.
\end{align*}
Therefore, we have confirmed that all the postulates in
Section~\ref{sec:palge} are fulfilled.
Applying the algebra and multiplication law, we can derive
the trilinear relations which characterize parafermionic statistics
\cite{GR52,OK82} as
\begin{align}
[\psi^{-},[\psi^{+},\psi^{-}]]=2\psi^{-}\psi^{+}\psi^{-}
 -\{\psi^{+},(\psi^{-})^{2}\}=\psi^{-},\\
[\psi^{+},[\psi^{-},\psi^{+}]]=2\psi^{+}\psi^{-}\psi^{+}
 -\{\psi^{-},(\psi^{+})^{2}\}=\psi^{+}.
\end{align}

\section{Second-Order Parasupersymmetric Quantum Systems}
\label{sec:2psqs}

We are now in a position to construct second-order parasupersymmetric
quantum systems by using the second-order parafermionic algebra
just derived in the previous section. From Eqs.~(\ref{eq:2pjo1}),
(\ref{eq:2pjo2}), and the multiplication law, the triple $(\bH,
\bQ^{-},\bQ^{+})$ in Eqs.~(\ref{eq:pfqs}) and (\ref{eq:pfsc}) for
$p=2$ is given by
\begin{align}
\bH&=\sum_{k=0}^{2}H_{k}\Pi_{k}\notag\\
&=H_{0}(\psi^{-})^{2}(\psi^{+})^{2}
 +H_{1}(\psi^{+}\psi^{-}-(\psi^{+})^{2}(\psi^{-})^{2})
 +H_{2}(\psi^{+})^{2}(\psi^{-})^{2},\\[5pt]
\bQ^{-}&=\sum_{k=0}^{2}Q_{k}^{-}\psi^{-}\Pi_{k}
 =Q_{1}^{-}(\psi^{-})^{2}\psi^{+}+Q_{2}^{-}\psi^{+}(\psi^{-})^{2},
\label{eq:psc2-}\\
\bQ^{+}&=\sum_{k=0}^{2}Q_{k}^{+}\Pi_{k}\psi^{+}
 =Q_{1}^{+}\psi^{-}(\psi^{+})^{2}+Q_{2}^{+}(\psi^{+})^{2}\psi^{-}.
\label{eq:psc2+}
\end{align}
We recall the fact that the above second-order parasupercharges
(\ref{eq:psc2-}) and (\ref{eq:psc2+}) already satisfy
the nilpotent condition (\ref{eq:pfsc1}) for $p=2$,
$(\bQ^{-})^{3}=(\bQ^{+})^{3}=0$, as we have generically shown in
Section~\ref{sec:psusy}.

Now that we have had the second-order parafermionic algebra and
multiplication law, we can reveal the true character of the
`discrete-type' symmetry mentioned in Section~\ref{sec:psusy}.
In the case of second-order, we can construct eight additional
charges $\bQ_{\{n\}}^{\pm}$ and $\bQ_{[n]}^{\pm}$ ($n=1,2$)
defined by Eq.~(\ref{eq:discc}). Substituting Eqs.~(\ref{eq:alg2b}),
(\ref{eq:2pjo1}), and (\ref{eq:2pjo2}) into Eqs.~(\ref{eq:discc1})
and using the multiplication law, we have
\begin{subequations}
\begin{align}
\bQ_{\{1\}}^{-}=-\bQ_{\{2\}}^{-}=-Q_{1}^{-}(\psi^{-})^{2}\psi^{+}
 +Q_{2}^{-}\psi^{+}(\psi^{-})^{2},\qquad
 \bQ_{[1]}^{-}=\bQ_{[2]}^{-}=\bQ^{-},\\
\bQ_{\{1\}}^{+}=-\bQ_{\{2\}}^{+}=Q_{1}^{+}\psi^{-}(\psi^{+})^{2}
 -Q_{2}^{+}(\psi^{+})^{2}\psi^{-},\qquad
 \bQ_{[1]}^{+}=\bQ_{[2]}^{+}=-\bQ^{+}.
\end{align}
\end{subequations}
Hence, there are essentially two new conserved charges
$\bQ_{\{1\}}^{\pm}$ obtained from $\bQ^{\pm}$ by changing
the relative signs between $Q_{1}^{\pm}$ and $Q_{2}^{\pm}$,
respectively. Apparently, they have $\bbZ_{2}$ structure and that
is the reason why we have called them `discrete-type' symmetries.
These conserved charges are identical with the ones first found
in Ref.~\cite{DV90a}.

From Eqs.~(\ref{eq:inter}) and (\ref{eq:parac}), the
commutativity (\ref{eq:pfsc2}) and the non-linear constraints
(\ref{eq:pfsc3}) for $p=2$
\begin{subequations}
\begin{align}
(\bQ^{-})^{2}\bQ^{+}+\bQ^{-}\bQ^{+}\bQ^{-}+\bQ^{+}(\bQ^{-})^{2}
 =C_{2}\bQ^{-}\bH,\\
(\bQ^{+})^{2}\bQ^{-}+\bQ^{+}\bQ^{-}\bQ^{+}+\bQ^{-}(\bQ^{+})^{2}
 =C_{2}\bH\bQ^{+},
\end{align}
\end{subequations}
hold if and only if the following conditions
\begin{align}
H_{0}Q_{1}^{-}=Q_{1}^{-}H_{1},\qquad
 H_{1}Q_{2}^{-}=Q_{2}^{-}H_{2},
\label{eq:2psc1}\\
Q_{1}^{-}Q_{2}^{-}Q_{2}^{+}+Q_{1}^{-}Q_{1}^{+}Q_{1}^{-}
 =C_{2}Q_{1}^{-}H_{1},
\label{eq:2psc2}\\
Q_{2}^{-}Q_{2}^{+}Q_{2}^{-}+Q_{1}^{+}Q_{1}^{-}Q_{2}^{-}
 =C_{2}Q_{2}^{-}H_{2},
\label{eq:2psc3}
\end{align}
and their `adjoint' relations
\begin{align}
Q_{1}^{+}H_{0}=H_{1}Q_{1}^{+},\qquad
 Q_{2}^{+}H_{1}=H_{2}Q_{2}^{+},
\label{eq:2psc1'}\\
Q_{1}^{+}Q_{1}^{-}Q_{1}^{+}+Q_{2}^{-}Q_{2}^{+}Q_{1}^{+}
 =C_{2}H_{1}Q_{1}^{+},
\label{eq:2psc2'}\\
Q_{2}^{+}Q_{1}^{+}Q_{1}^{-}+Q_{2}^{+}Q_{2}^{-}Q_{2}^{+}
 =C_{2}H_{2}Q_{2}^{+},
\label{eq:2psc3'}
\end{align}
are satisfied. We note that when a solution to Eq.~(\ref{eq:2psc2})
or (\ref{eq:2psc2'}) is given by
\begin{align}
C_{2}H_{1}=Q_{1}^{+}Q_{1}^{-}+Q_{2}^{-}Q_{2}^{+},
\label{eq:2psc4}
\end{align}
the conditions (\ref{eq:2psc3}) and (\ref{eq:2psc3'}) become
identical with the second intertwining relations in
Eqs.~(\ref{eq:2psc1}) and (\ref{eq:2psc1'}), respectively.
Thus, in this case it is sufficient to solve the following operator
identities
\begin{align}
(C_{2}H_{0}-Q_{1}^{-}Q_{1}^{+})Q_{1}^{-}
 =Q_{1}^{-}Q_{2}^{-}Q_{2}^{+},
\label{eq:2psc5}\\
Q_{2}^{-}(C_{2}H_{2}-Q_{2}^{+}Q_{2}^{-})
 =Q_{1}^{+}Q_{1}^{-}Q_{2}^{-}.
\label{eq:2psc6}
\end{align}
In general, we do not need to solve the `adjoint' conditions.

For the second-order case, we have one new quasi-parasupersymmetry,
namely, that of order $(2,2)$. The conditions are given by
Eqs.~(\ref{eq:2psc1})--(\ref{eq:2psc3'}), but the first-order
intertwining relations (\ref{eq:2psc1}) and (\ref{eq:2psc1'}) are
replaced by the second-order intertwining relations
\begin{align}
H_{0}Q_{1}^{-}Q_{2}^{-}=Q_{1}^{-}Q_{2}^{-}H_{2},\qquad
Q_{2}^{+}Q_{1}^{+}H_{0}=H_{2}Q_{2}^{+}Q_{1}^{+}.
\label{eq:2qpsc1}
\end{align}
In the followings, we will show three different representations
for the system $(H_{k},Q_{k}^{\pm})$ which satisfies the condition
(\ref{eq:2psc1})--(\ref{eq:2psc3'}) for second-order
parasupersymmetry.

\subsection{One-Variable Representation}
\label{ssec:exam1}

First, we shall realize a second-order parasupersymmetric quantum
system of one degree of freedom. Let us put $C_{2}=4$ and
\begin{align}
H_{k}=-\frac{1}{2}\del^{2}+V_{k}(q),
 \qquad Q_{k}^{\pm}=\pm\del+W_{k}(q),
\label{eq:1vrep}
\end{align}
where $\del=\rmd/\rmd q$. Substituting Eq.~(\ref{eq:1vrep}) into
Eq.~(\ref{eq:2psc2}), we find that the condition (\ref{eq:2psc2})
is satisfied if and only if
\begin{align}
4V_{1}=W'_{1}+W_{1}^{2}-W'_{2}+W_{2}^{2}.
\end{align}
Hence, the general solution to (\ref{eq:2psc2}) is given by
Eq.~(\ref{eq:2psc4}). In this case we have
\begin{align}
C_{2}H_{0}-Q_{1}^{-}Q_{1}^{+}=-\del^{2}+4V_{0}+W'_{1}-W_{1}^{2}
 \equiv -\del^{2}+2\bar{V}_{0}.
\end{align}
Then, the condition (\ref{eq:2psc5}) is equivalent to the following
set of equations:
\begin{align}
2\bar{V}_{0}=-2W'_{1}-W'_{2}+W_{2}^{2},\\
2\bar{V}'_{0}+2W_{1}\bar{V}_{0}=-W''_{1}-W_{1}W'_{2}+W_{1}W_{2}^{2}.
\end{align}
From them we immediately obtain
\begin{align}
-W'_{2}+W_{2}^{2}=W'_{1}+W_{1}^{2}+4C,\qquad 2\bar{V}_{0}
 =-W'_{1}+W_{1}^{2}+4C,
\label{eq:pfc1}
\end{align}
where $C$ is a constant. In terms of $H_{i}$ and $Q_{i}^{\pm}$
they are expressed as
\begin{align}
Q_{2}^{-}Q_{2}^{+}=Q_{1}^{+}Q_{1}^{-}+4C,\qquad
 C_{2}H_{0}=2Q_{1}^{-}Q_{1}^{+}+4C.
\label{eq:pfc1'}
\end{align}
Substituting the first formula in Eq.~(\ref{eq:pfc1'}) into
the condition Eq.~(\ref{eq:2psc6}), we obtain
\begin{align}
C_{2}Q_{2}^{-}H_{2}=2Q_{2}^{-}Q_{2}^{+}Q_{2}^{-}-4CQ_{2}^{-}.
\end{align}
This identity holds if and only if
\begin{align}
2V_{2}=W'_{2}+W_{2}^{2}-2C.
\end{align}
Hence, we finally obtain
\begin{align}
H_{0}=\frac{1}{2}Q_{1}^{-}Q_{1}^{+}+C,\quad H_{1}=\frac{1}{2}
 Q_{1}^{+}Q_{1}^{-}+C=\frac{1}{2}Q_{2}^{-}Q_{2}^{+}-C,\quad
 H_{2}=\frac{1}{2}Q_{2}^{+}Q_{2}^{-}-C.
\label{eq:pfc2}
\end{align}
This second-order parasupersymmetric system is essentially the same
as the one first constructed by Rubakov and Spiridonov in
Ref.~\cite{RS88}. We also note that if we put
\begin{align}
W_{1}(q)=W(q)+\frac{E(q)}{2},\qquad
 W_{2}(q)=W(q)-\frac{E(q)}{2},
\end{align}
the Hamiltonians $H_{0}$ and $H_{2}$ are intertwined by
\begin{align}
Q_{2}^{+}Q_{1}^{+}=\left(\frac{\rmd}{\rmd q}+W(q)
 -\frac{E(q)}{2}\right)\left(\frac{\rmd}{\rmd q}+W(q)
 +\frac{E(q)}{2}\right),
\end{align}
which is exactly the component of type A $2$-fold supercharge,
and thus they constitute a pair of type A $2$-fold
supersymmetry \cite{Ta03a,AST01a}.

\subsection{Two-Variable Representation I}
\label{ssec:exam2}

If we consider the fact that each parasupercharge of order 2
has two independent components, it would be natural to expect
that a second-order parasupersymmetry can be realized in
quantum systems of two degrees of freedom.
For this purpose, we put $C_{2}=2$ and
\begin{align}
H_{k}=-\frac{1}{2}\del_{1}^{2}-\frac{1}{2}\del_{2}^{2}
 +V_{k}(q_{1},q_{2}),\qquad Q_{k}^{\pm}=\pm\del_{k}
 +W_{k}(q_{1},q_{2}),
\label{eq:2vrep1}
\end{align}
where $\del_{k}=\del/\del q_{k}$. Substituting Eq.~(\ref{eq:2vrep1})
into Eq.~(\ref{eq:2psc2}), we find that the condition (\ref{eq:2psc2})
is satisfied if and only if
\begin{align}
2V_{1}=(\del_{1}W_{1})+W_{1}^{2}-(\del_{2}W_{2})+W_{2}^{2},
\end{align}
where $(\del_{i}f)=\del f/\del q_{i}$ and thus the general
solution is again given by Eq.~(\ref{eq:2psc4}). In this case,
\begin{align}
C_{2}H_{0}-Q_{1}^{-}Q_{1}^{+}=-\del_{2}^{2}+2V_{0}+(\del_{1}W_{1})
 -W_{1}^{2}\equiv -\del_{2}^{2}+2\bar{V}_{0},
\end{align}
and the condition (\ref{eq:2psc5}) is equivalent to the following
set of equations:
\begin{align}
(\del_{2}W_{1})=0,\\
2\bar{V}_{0}=-(\del_{2}W_{2})+W_{2}^{2},\\
2(\del_{1}\bar{V}_{0})+2W_{1}\bar{V}_{0}=-W_{1}(\del_{2}W_{2})
 +W_{1}W_{2}^{2}.
\end{align}
From them we have
\begin{align}
2(\del_{1}\bar{V}_{0})=-(\del_{1}\del_{2}W_{2})
 +(\del_{1}W_{2}^{2})=0.
\end{align}
On the other hand,
\begin{align}
C_{2}H_{2}-Q_{2}^{+}Q_{2}^{-}=-\del_{1}^{2}+2V_{2}-(\del_{2}W_{2})
 -W_{2}^{2}\equiv -\del_{1}^{2}+2\bar{V}_{2},
\end{align}
and the condition (\ref{eq:2psc6}) is equivalent to the following
set of equations:
\begin{align}
(\del_{1}W_{2})=0,\\
2\bar{V}_{2}=(\del_{1}W_{1})+W_{1}^{2},\\
-2(\del_{2}\bar{V}_{2})+2W_{2}\bar{V}_{2}=(\del_{1}W_{1})W_{2}
 +W_{1}^{2}W_{2}.
\end{align}
From them we have
\begin{align}
2(\del_{2}\bar{V}_{2})=(\del_{2}\del_{1}W_{1})
 +(\del_{2}W_{1}^{2})=0.
\end{align}
Summarizing the above results, we obtain for the parasupercharges
\begin{align}
Q_{k}^{\pm}=\pm\del_{k}+W_{k}(q_{k})\quad (k=1,2),
\end{align}
and for the Hamiltonians
\begin{align}
H_{0}&=\frac{1}{2}(Q_{1}^{-}Q_{1}^{+}+Q_{2}^{-}Q_{2}^{+})\notag\\
&=-\frac{1}{2}\del_{1}^{2}-\frac{1}{2}\del_{2}^{2}+\frac{1}{2}
 \left(-W'_{1}(q_{1})+W_{1}(q_{1})^{2}-W'_{2}(q_{2})
 +W_{2}(q_{2})^{2}\right),\\
H_{1}&=\frac{1}{2}(Q_{1}^{+}Q_{1}^{-}+Q_{2}^{-}Q_{2}^{+})\notag\\
&=-\frac{1}{2}\del_{1}^{2}-\frac{1}{2}\del_{2}^{2}+\frac{1}{2}
 \left(W'_{1}(q_{1})+W_{1}(q_{1})^{2}-W'_{2}(q_{2})
 +W_{2}(q_{2})^{2}\right),\\
H_{2}&=\frac{1}{2}(Q_{1}^{+}Q_{1}^{-}+Q_{2}^{+}Q_{2}^{-})\notag\\
&=-\frac{1}{2}\del_{1}^{2}-\frac{1}{2}\del_{2}^{2}+\frac{1}{2}
 \left(W'_{1}(q_{1})+W_{1}(q_{1})^{2}+W'_{2}(q_{2})
 +W_{2}(q_{2})^{2}\right),
\end{align}
where $W_{k}$ ($k=1,2$) are arbitrary differentiable functions.
Hence, the system consists of two independent particles
subject to the different external potentials.

\subsection{Two-Variable Representation II}
\label{ssec:exam3}

To construct a parasupersymmetric interacting two-body system,
we put $C_{2}=4$ and
\begin{subequations}
\label{eq:2vrep2}
\begin{align}
H_{k}&=-\frac{1}{2}\del_{1}^{2}-\frac{1}{2}\del_{2}^{2}
 +V_{k}(q_{1},q_{2}),\\
Q_{1}^{\pm}&=\pm\del_{1}\mp\del_{2}+W_{1}(q_{1},q_{2}),\\
Q_{2}^{\pm}&=\pm\del_{1}\pm\del_{2}+W_{2}(q_{1},q_{2}).
\end{align}
\end{subequations}
As in the previous two examples, the general solution to
Eq.~(\ref{eq:2psc2}) is again given by Eq.~(\ref{eq:2psc4}).
In this case we have
\begin{align}
C_{2}H_{0}-Q_{1}^{-}Q_{1}^{+}&=-\del_{1}^{2}-2\del_{1}\del_{2}
 -\del_{2}^{2}+4V_{0}+(\del_{1}W_{1})-(\del_{2}W_{1})-W_{1}^{2}
 \notag\\
&\equiv -\del_{1}^{2}-2\del_{1}\del_{2}-\del_{2}^{2}
 +2\bar{V}_{0},\\
C_{2}H_{2}-Q_{2}^{+}Q_{2}^{-}&=-\del_{1}^{2}+2\del_{1}\del_{2}
 -\del_{2}^{2}+4V_{2}-(\del_{1}W_{2})-(\del_{2}W_{2})-W_{2}^{2}
 \notag\\
&\equiv -\del_{1}^{2}+2\del_{1}\del_{2}-\del_{2}^{2}
 +2\bar{V}_{2}.
\end{align}
Then, it is easy to check that the condition (\ref{eq:2psc5})
is satisfied if and only if
\begin{align}
(\del_{1}W_{1})+(\del_{2}W_{1})=0,
\label{eq:2psc51}\\
2\bar{V}_{0}=-(\del_{1}W_{2})-(\del_{2}W_{2})+W_{2}^{2},\\
-(\del_{1}^{2}W_{2})+(\del_{1}W_{2}^{2})+(\del_{2}^{2}
 W_{2})-(\del_{2}W_{2}^{2})=0.
\label{eq:2psc53}
\end{align}
Similarly, the condition (\ref{eq:2psc6}) is satisfied if and
only if
\begin{align}
(\del_{1}W_{2})-(\del_{2}W_{2})=0,
\label{eq:2psc61}\\
2\bar{V}_{2}=(\del_{1}W_{1})-(\del_{2}W_{1})+W_{1}^{2},\\
(\del_{1}^{2}W_{1})+(\del_{1}W_{1}^{2})-(\del_{2}^{2}W_{1})
 +(\del_{2}W_{1}^{2})=0.
\label{eq:2psc63}
\end{align}
We note that Eqs.~(\ref{eq:2psc51}) and (\ref{eq:2psc61})
imply Eqs.~(\ref{eq:2psc63}) and (\ref{eq:2psc53}), respectively.
Summarizing the results, we obtain for the parasupercharges
\begin{align}
Q_{1}^{\pm}&=\pm\del_{1}\mp\del_{2}+W_{1}(q_{1}-q_{2}),\\
Q_{2}^{\pm}&=\pm\del_{1}\pm\del_{2}+W_{2}(q_{1}+q_{2}),
\end{align}
and for the Hamiltonians
\begin{align}
H_{0}&=\frac{1}{4}(Q_{1}^{-}Q_{1}^{+}+Q_{2}^{-}Q_{2}^{+})\notag\\
&=-\frac{1}{2}\del_{1}^{2}-\frac{1}{2}\del_{2}^{2}+\frac{1}{4}
 \left(-W'_{1}(q_{-})+W_{1}(q_{-})^{2}
 -W'_{2}(q_{+})+W_{2}(q_{+})^{2}\right),\\
H_{1}&=\frac{1}{4}(Q_{1}^{+}Q_{1}^{-}+Q_{2}^{-}Q_{2}^{+})\notag\\
&=-\frac{1}{2}\del_{1}^{2}-\frac{1}{2}\del_{2}^{2}+\frac{1}{4}
 \left(W'_{1}(q_{-})+W_{1}(q_{-})^{2}
 -W'_{2}(q_{+})+W_{2}(q_{+})^{2}\right),\\
H_{2}&=\frac{1}{4}(Q_{1}^{+}Q_{1}^{-}+Q_{2}^{+}Q_{2}^{-})\notag\\
&=-\frac{1}{2}\del_{1}^{2}-\frac{1}{2}\del_{2}^{2}+\frac{1}{4}
 \left(W'_{1}(q_{-})+W_{1}(q_{-})^{2}
 +W'_{2}(q_{+})+W_{2}(q_{+})^{2}\right),
\end{align}
where $q_{\pm}=q_{1}\pm q_{2}$ and $W_{k}$ ($k=1,2$) are arbitrary
differentiable functions.
Hence, the potential terms of the parasupersymmetric system in
this case consist of the interactions depending only on the
relative coordinate $q_{1}-q_{2}$ and the external fields
acting on the center-of-mass coordinate $q_{1}+q_{2}$.

\subsection{Summary of the three examples}

In contrast to supersymmetric quantum systems, it is well-known
that parasupersymmetric ones do not necessarily have non-negative
spectrum and do not generically admit a generalization of the
Witten index. These facts led to consider some additional
non-linear relations among $\bH$ and $\bQ^{\pm}$, in particular,
some cases where the system $\bH$ can be expressed as
a non-negative function of $\bQ^{\pm}$ in closed form, see
e.g. Refs.~\cite{KMR93a,Mo96a,Mo97}. In the latter cases, since
the operator $\bH$ is of parafermionic degree 0, each term in
a function of $\bQ^{\pm}$ should contain the same number of
$\bQ^{-}$ and $\bQ^{+}$. In the case of second-order parasupersymmetry,
non-trivial such monomials of the lowest degree in $\bQ^{\pm}$ are
as follows:
\begin{align}
\bQ^{-}\bQ^{+}=Q_{1}^{-}Q_{1}^{+}\Pi_{0}+Q_{2}^{-}Q_{2}^{+}\Pi_{1},
\label{eq:mono1}\\
\bQ^{+}\bQ^{-}=Q_{1}^{+}Q_{1}^{-}\Pi_{1}+Q_{2}^{+}Q_{2}^{-}\Pi_{2}.
\end{align}
Similarly, those of the next-to-lowest degree in $\bQ^{\pm}$ which
cannot be obtained by a function of $\bQ^{-}\bQ^{+}$ and $\bQ^{+}\bQ^{-}$
are as follows:
\begin{align}
(\bQ^{-})^{2}(\bQ^{+})^{2}&=Q_{1}^{-}Q_{2}^{-}Q_{2}^{+}Q_{1}^{+}\Pi_{0},\\
(\bQ^{+})^{2}(\bQ^{-})^{2}&=Q_{2}^{+}Q_{1}^{+}Q_{1}^{-}Q_{2}^{-}\Pi_{2}.
\label{eq:mono4}
\end{align}
Owing to the nilpotency $(\bQ^{\pm})^{3}=0$, every function of
$\bQ^{\pm}$ which has zeroth parafermionic degree can be expressed,
in principle, as a function of the monomials
(\ref{eq:mono1})--(\ref{eq:mono4}).

Let us now consider the first example in Section~\ref{ssec:exam1}.
Using the first relation in Eqs.~(\ref{eq:pfc1'}) and
(\ref{eq:pfc2}), we obtain
\begin{align}
\bQ^{-}\bQ^{+}&=2(H_{0}-C)\Pi_{0}+2(H_{1}+C)\Pi_{1},\\
\bQ^{+}\bQ^{-}&=2(H_{1}-C)\Pi_{1}+2(H_{2}+C)\Pi_{2},\\
(\bQ^{-})^{2}(\bQ^{+})^{2}&=4(H_{0}^{2}-C^{2})\Pi_{0},\\
(\bQ^{+})^{2}(\bQ^{-})^{2}&=4(H_{2}^{2}-C^{2})\Pi_{2}.
\end{align}
Hence, we can easily find a non-linear relation
\begin{align}
(\bQ^{-})^{2}(\bQ^{+})^{2}+\bQ^{\pm}(\bQ^{\mp})^{2}\bQ^{\pm}
 +(\bQ^{+})^{2}(\bQ^{-})^{2}=4(\bH^{2}-C^{2}).
\end{align}
It is interesting to note that this non-linear relation can be
regarded as a generalization of 2-fold superalgebra. Indeed, if we
restrict the linear space $\fF\times\sV_{2}$ on which the system $\bH$
acts to $\fF\times(\sV_{2}^{(0)}\dotplus\sV_{2}^{(2)})$ (cf. the
definition between Eqs.~(\ref{eq:defpfs}) and (\ref{eq:defadv})),
we have
\begin{align}
\{(\bQ^{-})^{2},(\bQ^{+})^{2}\}=4(\bH^{2}-C^{2})
 \bigr|_{\fF\times(\sV_{2}^{(0)}\dotplus\sV_{2}^{(2)})}.
\end{align}
This, together with the trivial (anti-)commutation relations
\begin{align}
\{(\bQ^{-})^{2},(\bQ^{-})^{2}\}=\{(\bQ^{+})^{2},(\bQ^{+})^{2}\}
 =[(\bQ^{\pm})^{2},\bH]=0,
\end{align}
constitutes a type of 2-fold superalgebra in the sector
$\fF\times(\sV_{2}^{(0)}\dotplus\sV_{2}^{(2)})$.

In the second and third examples in Sections~\ref{ssec:exam2} and
\ref{ssec:exam3}, the parasupersymmetric system $\bH$ is given by
\begin{align}
C_{2}\bH=&\, (Q_{1}^{-}Q_{1}^{+}+Q_{2}^{-}Q_{2}^{+})\Pi_{0}
 +(Q_{1}^{+}Q_{1}^{-}+Q_{2}^{-}Q_{2}^{+})\Pi_{1}\notag\\
&\, +(Q_{1}^{+}Q_{1}^{-}+Q_{2}^{+}Q_{2}^{-})\Pi_{2}.
\end{align}
Here we can observe a different situation from the previous one-body
case. Since $\Pi_{0}$-component of the system $\bH$ has the term
$Q_{2}^{-}Q_{2}^{+}$, $\Pi_{0}$-component of $\bH^{n}$ always includes
the term $(Q_{2}^{-}Q_{2}^{+})^{n}$. However, the latter term in
$\Pi_{0}$-component cannot be reproduced by any function of the
operators in Eqs.~(\ref{eq:mono1})--(\ref{eq:mono4}). The same is
true for the term $Q_{1}^{+}Q_{1}^{-}$ in $\Pi_{2}$-component.
Hence, in the second and third cases we cannot express any function
of $\bH$ in terms of $\bQ^{\pm}$.

Finally, we note that for all the choices (\ref{eq:1vrep}),
(\ref{eq:2vrep1}), and (\ref{eq:2vrep2}) in the three examples,
quasi-parasupersymmetry of order $(2,2)$ does not produce any new
result over parasupersymmetry of order $2$. The reason is that in
all the three cases the solution to the non-linear constraint
(\ref{eq:2psc2}) is given by Eq.~(\ref{eq:2psc4}). In this case,
as we have already mentioned previously, the other non-linear
constraint (\ref{eq:2psc3}) is identical with the first-order
intertwining relation between $H_{1}$ and $H_{2}$, namely,
the second relation in Eqs.~(\ref{eq:2psc1}) and (\ref{eq:2psc1'}).
As a result, the condition (\ref{eq:2qpsc1}) for
quasi-parasupersymmetry is equivalent to
\begin{align}
H_{0}Q_{1}^{-}Q_{2}^{-}=Q_{1}^{-}H_{1}Q_{2}^{-},\quad
 Q_{2}^{+}Q_{1}^{+}H_{0}=Q_{2}^{+}H_{1}Q_{1}^{+},
\end{align}
which is very close to the first relation in Eqs.~(\ref{eq:2psc1})
and (\ref{eq:2psc1'}). Hence, in most of second-order cases,
quasi-parasupersymmetry would be identical to parasupersymmetry.

\section{Discussion and Summary}
\label{sec:discus}

In this article, we have proposed the systematic construction
of parafermionic algebra and parasupersymmetric quantum systems.
Assuming relatively small number of assumptions, namely,
Eqs.~(\ref{eq:postu}), (\ref{eq:mult1}), and (\ref{eq:mult3}) in
addition to the nilpotency (\ref{eq:nilpo}), we have shown that
we can systematically construct the full set of parafermionic
algebra and multiplication law, at least for second order, which
are totally independent of any specific representation of
parafermionic operators.

With the aid of the parafermionic algebra, we have formulated
in the generic way parasupersymmetric quantum systems.
Remarkably, we have shown that all the parasupersymmetric
conditions, namely, the commutation relations (\ref{eq:pfsc2})
and the non-linear relations (\ref{eq:pfsc3}) can be expressed
in closed form in terms of the component scalar Hamiltonians
$H_{k}$ and the component parasupercharges $Q_{k}^{\pm}$.
From those expressions we have found that every pair of the
component Hamiltonians possesses $\cN$-fold supersymmetry with
$\cN\leq q$ and thus has isospectral property and weak
quasi-solvability. This means in particular that parasupersymmetric
quantum field theory (for some attempts of construction, see, e.g.,
Refs.~\cite{FV91,BD93b}), if exists, should have characteristic
features analogous to those of $\cN$-fold supersymmetry, as in the
case of weak supersymmetry (cf. Section~\ref{sec:intro}).
The fact that the parasupersymmetric condition is too strong
has naturally led us to the generalization of parasupersymmetry,
which we have called quasi-parasupersymmetry.

We have investigated second-order parasupersymmetric quantum
systems in detail and exhibited the three simple examples of such
systems, the first one is essentially equivalent to the original
one-body system in Ref.~\cite{RS88} and the others are two-body
systems in which two independent supersymmetries are folded. In
particular, we have shown that the first model admits a generalization
of $2$-fold superalgebra. In all the three models, quasi-parasupersymmetry
is equivalent to parasupersymmetry.

Constructions of higher-order parafermionic algebra and
(quasi-)parasupersymmetric quantum systems are straightforward.
An extensive study of higher-order cases would be reported
in a subsequent publication~\cite{Ta07c}.

In this article, we have restricted ourselves to ordinary
Schr\"odinger operators (\ref{eq:Schro}) as representations
of parasuperalgebra, there are of course other possible
applications in various areas of physics and mathematical
science. One of them is the application to quantum systems
with position-dependent mass (PDM) described generically by
von Roos operators~\cite{vR83}. It would be straightforward
since $\cN$-fold supersymmetry has been already successfully
formulated also for PDM quantum systems~\cite{Ta06a}. For
instance, we can generalize the second example of second-order
parasupersymmetry in Section~\ref{ssec:exam2} to PDM two-body
quantum systems by choosing the component parasupercharges as
\begin{align}
Q_{k}^{\pm}=m_{k}(q_{k})^{-\frac{1}{2}}\left(\pm\frac{\del}{
 \del q_{k}}+W_{k}(q_{k})-\frac{m'_{k}(q_{k})}{4m_{k}(q_{k})}
 \right)\quad (k=1,2).
\end{align}

As we have mentioned in Section~\ref{sec:intro}, there have
been various formulations of parasupersymmetry. Thus, it would
be interesting and important to examine the relationship among
them including the present formulation.

A generalization of the present formulation to several
parafermionic variables $(\psi_{1}^{-},\dots,\psi_{n}^{-},
\psi_{1}^{+},\dots,\psi_{n}^{+})$ is a challenging problem.
The nilpotency (\ref{eq:nilpo}) can be trivially generalized as
$(\psi_{i}^{-})^{p+1}=(\psi_{i}^{+})^{p+1}=0$ for all
$i=1,\dots,n$. A naive generalization of the fermionic
anti-commutation relation $\{\psi_{i}^{-},\psi_{j}^{+}\}
=\delta_{i,j}I$, which reduces to Eq.~(\ref{eq:postu}) when
$n=1$ may be
\begin{align}
\{\psi_{i}^{-},\psi_{j}^{+}\}+\{(\psi_{i}^{-})^{2},
 (\psi_{j}^{+})^{2}\}+\dots+\{(\psi_{i}^{-})^{p},
 (\psi_{j}^{+})^{p}\}=p\delta_{i,j}I.
\end{align}
It would be interesting to examine whether this kind of
generalization to several parafermionic variables works well.

%\section*{Acknowledgments}
\begin{acknowledgments}%REVTEX4
%\begin{ack}%ELSART
%\ack%IOPART
 We would like to thank C.~Quesne for the valuable discussion
 in the earliest stage of the project.
 This work was partially supported by the National Cheng-Kung
 University under the grant No. OUA:95-3-2-071.
\end{acknowledgments}%REVTEX4
%\end{ack}%ELSART

%\appendix

%\section*{References}%IOPART

\bibliography{refsels}%BIB-FILE
\bibliographystyle{npb}%BST-FILE
%\begin{thebibliography}{99}
%\def\J#1#2#3#4{{\sl #1} {\bf #2} (#3) #4}

%\bibitem{}
%Author 1, Author 2 and Author 3,
%\J{Journal}{Volume}{Year}{Page}.

%\end{thebibliography}

\end{document}